\documentclass[a4paper, 10pt]{amsproc}

\usepackage{defs}
\usepackage{amsaddr, orcidlink}

\title[Algorithmic construction of Lyapunov functions]{On the algorithmic construction of Lyapunov functions for continuous vector fields} 

\author{Raavi Gupta\,\orcidlink{0009-0003-7585-5820}, Sameep Chattopadhyay\,\orcidlink{0009-0006-4474-8266}}
\address{%
	\faGroup\ Department of Electrical Engineering\\
	\faUniversity\ Indian Institute of Technology Bombay\\
	\faMapMarker\ Powai, Mumbai 400076, India
}

\author[R. Gupta, S. Chattopadhyay, P. Paruchuri, and D. Chatterjee]{Pradyumna Paruchuri\,\orcidlink{0000-0002-8598-5069}, and Debasish Chatterjee\,\orcidlink{0000-0002-1718-653X}}
\address{%
	\faGroup\ Systems \& Control Engineering\\
	\faUniversity\ Indian Institute of Technology Bombay\\
	\faMapMarker\ Powai, Mumbai 400076, India
}
\thanks{%
	\faHome\ (DC) \url{https://www.sc.iitb.ac.in/~chatterjee}\\
	\indent	\faEnvelope\ \texttt{raavi02g@gmail.com, sameep.ch.2002@gmail.com, pradyumnaparuchuri@gmail.com, dchatter@iitb.ac.in}
}

\thanks{D.\ Chatterjee acknowledges partial support of the SERB MATRICS grant MTR/2022/000656 from the Govt.\ of India.}

\date{\DTMnow}

\begin{document}

\begin{abstract}
	This article presents a novel numerically tractable technique for synthesizing Lyapunov functions for equilibria of nonlinear vector fields. In broad strokes, corresponding to an isolated equilibrium point of a given vector field, a selection is made of a compact neighborhood of the equilibrium and a dictionary of functions in which a Lyapunov function is expected to lie. Then an algorithmic procedure based on the recent work \cite{ref:DasAraCheCha-22} is deployed on the preceding neighborhood-dictionary pair and charged with the task of finding a function satisfying a compact family of inequalities that defines the behavior of a Lyapunov function on the selected neighborhood. The technique applies to continuous nonlinear vector fields without special algebraic structures and does not even require their analytical expressions to proceed. Several numerical examples are presented to illustrate our results.
\end{abstract}

\maketitle

\section{Background, setting of the problem, and our contributions}
\label{s:intro}

\subsection{Nonlinear systems and their equilibria}

Nonlinear systems arise in a vast plethora of engineering applications ranging from classical mechanics through mathematical finance and are one of the key driving engines behind the success of today's popular optimization algorithms. A nonlinear system is described by an ordinary differential equation
\begin{equation}
	\label{e:nonlinear system}
	\dot{\state}(t) = \vecfld\bigl(\state(t)\bigr),
\end{equation}
where \(\state(t)\in\R[\sysDim]\) is the vector of states of the system defined on a nonempty open set \(\domain\subset \R[\sysDim]\), and \(\vecfld:\domain\lra\R[\sysDim]\) is a \embf{vector field} that specifies at each state \(y\in\domain\) a vector \(\vecfld(y)\) describing both the instantaneous direction of motion and the magnitude of the velocity of motion in that direction. It is assumed that \(\vecfld\) is a Lipschitz continuous map. A nonlinear system is written frequently in the form of a so-called \emph{Cauchy problem}, which specifies a boundary condition such as \(\state(0) = \ol\state\); given such a boundary condition, the system \eqref{e:nonlinear system} can be solved uniquely in a small enough neighborhood of \(t = 0\); in this context, an absolutely continuous function \(\loro{-\eps}{\eps}\ni t\mapsto \soln(t, \ol\state)\in\R[\sysDim]\) for some \(\eps > 0\) such that \(\soln(0, \ol\state) = \ol\state\) and \(\pdv{\soln}{t}(s, \ol\state) = \vecfld\bigl( \soln(s, \ol\state) \bigr)\) for all \(s\in\loro{-\eps}{\eps}\) is the \embf{solution} to the nonlinear system \eqref{e:nonlinear system}, and the map \((t, x)\mapsto \soln(t, x)\) is its \embf{flow}.

An \embf{equilibrium point} of \eqref{e:nonlinear system} is a solution to the equation
\begin{equation}
    \label{e:equilibrium point}
	f(y) = 0\quad \text{for }y\in\R[\sysDim].
\end{equation}
A nonlinear system may admit no equilibrium point, and there may be finitely many equilibria or a continuum of them. In the early stages of the development of the field of dynamical systems, it was realized that the qualitative theory of equilibrium points plays a central role in both theory and applications, and the key concepts of stability of such points were developed in the late \(19\)-th century in A.\ M.\ Lyapunov's dissertation. After decades of mild obscurity, the concepts found applications in several engineering disciplines in the 1940s, and continue to be of central significance in modern day cybernetics.

\subsection{Lyapunov's indirect method}
\label{s:intro:indirect method}

Suppose that \(0\in\R[\sysDim]\) is an isolated equilibrium point of the vector field \(\vecfld\).\footnote{The selection of \(0\) is without loss of generality.}  Here are two key definitions (see e.g., \cite[Chapter V]{ref:Vid-02}) concerning stability of \(0\):
\begin{enumerate}[label=\textup{(S\arabic*)}, leftmargin=*, align=left]
	\item \label{stab:lyapstable} The equilibrium point \(0\) is \embf{Lyapunov stable} for every \(\eps > 0\) there exists \(\delta > 0\) such that \(\norm{\state(0)} < \delta\) implies that \(\norm{\soln\bigl(t, \state(0)\bigr)} < \eps\) for all \(t\ge 0\). Intuitively, \(0\) is Lyapunov stable if trajectories initialized close to \(0\) remain close to \(0\) for all time, and the extent of `closeness' can be made arbitrarily small (\(\eps\) is arbitrary).
	\item \label{stab:asystable} The equilibrium point \(0\) is \embf{asymptotically stable} if it is Lyapunov stable and \emph{asymptotically attractive}, i.e., for every \(\eps > 0\) there exist \(r > 0\) and \(T > 0\) such that \(\norm{\state(0)} < r\) and \(t > T\) implies \(\norm{\soln\bigl(t, \state(0)\bigr)} < \eps\).
\end{enumerate}
Both these definitions cater to local properties, and we adhere to the preceding premise throughout the sequel.

Lyapunov's indirect method (for detailed and illuminating treatments, see, e.g., \cite[Chapter 5]{ref:Vid-02}, \cite[Chapters V, VIII]{ref:BhaSze-70}, \cite[Chapter V]{ref:Hah-67}) provides a mechanism to test the nature of stability of an equilibrium point. The starting point in the case of Lyapunov stability, for instance, is a function \(\domain\ni y\mapsto \lyapfn(y)\in\R[]\)  (and \(\nbhd\subset\domain\) being a neighborhood of the origin \(0\in\R[\sysDim]\)) satisfying
\begin{equation}
	\label{e:Lyapunov function}
	\begin{aligned}
		& \lyapfn(0) = 0,\quad\text{and}\\
		& \begin{dcases}
			\lyapfn(y) > 0 & \text{for all }y\in\nbhd\setmin\set{0},\\
            \inprod{\pdv{\lyapfn}{x}(y)}{\vecfld(y)} \le 0 & \text{for all } y\in\nbhd,
		\end{dcases}
	\end{aligned}
\end{equation}
and Lyapunov's theorem asserts that if such a function \(\lyapfn\) exists, then \(0\) is Lyapunov stable. Notice that theorem substitutes the verification of a temporal property in \ref{stab:lyapstable} of the solution to the differential equation from arbitrary initial conditions sufficiently close to \(0\), with the verification of certain spatial properties of \(\lyapfn\) in \eqref{e:Lyapunov function}; in particular, one does not have to \emph{solve} the differential equation to assess stability of \(0\) via Lyapunov's indirect method. Naturally and on the one hand, finding such \embf{Lyapunov functions} \(\lyapfn\) is of tremendous utility, but on the other hand, the synthesis of Lyapunov functions is difficult since algorithmic and numerically tractable recipes for finding such functions are rare.\footnote{The situation is indeed rarified to the point of there being the folklore that arriving at suitable Lyapunov functions is more an art than science.} In certain mechanical systems one may write down with relative ease certain `energy'-based Lyapunov functions inspired by the physics of such problems, but a general recipe has proved to be elusive.

\subsection{Our contributions}

The central objective of this article is to fill the aforementioned lacuna: we provide a relatively general algorithmic recipe for constructing Lyapunov functions corresponding to isolated equilibria of dynamical systems. Our contributions are listed below, in the course of which we adopt comparative rhetoric to delineate them succinctly:
\begin{enumerate}[label=\textup{(\Alph*)}, align=left, widest=B, leftmargin=*]
	\item \label{contrib:} An algorithmic search for Lyapunov functions for a given equilibrium point of a continuous vector field is difficult: for one, there is no canonical parametrization of continuously differentiable positive definite functions to facilitate the construction of algorithms. Attempts have, therefore, been directed toward treating special families of vector fields and classes of admissible functions. The state of the art can be broadly classified into the following two regimes:
		\begin{itemize}[label=\(\circ\), leftmargin=*]
			\item \label{contrib:history} Algorithmic constructions of Lyapunov functions for \emph{linear vector fields} is easy \cite[\S15.10]{ref:Ber-18}. Indeed, if \(\vecfld(y) = A y\) for some matrix \(A\in\R[\sysDim\times\sysDim]\) with \(\rank A = \sysDim\),\footnote{Without the rank condition we have nontrivial subspaces of equilibria and such a situation is not within the scope of this article.} we know that quadratic Lyapunov functions of the form \(\R[\sysDim]\ni y\mapsto \lyapfn(y) = \inprod{y}{Py}\) for a suitable symmetric and positive definite matrix \(P\in\R[\sysDim\times\sysDim]\) suffice: the \embf{Lyapunov equation}
				\begin{equation}
					\label{e:Lyapunov equation}
					\trnsp{A} P + P A = -Q
				\end{equation}
				in the pair \((P, Q)\) is the centerpiece in this theory, and
				\begin{itemize}[label=\(\triangleright\), leftmargin=*]
					\item if there exists a symmetric and non-negative definite \(Q\) and a symmetric and positive definite \(P\) satisfying \eqref{e:Lyapunov equation}, then \(0\) is Lyapunov stable, and
					\item if there exist symmetric and positive definite matrices \(P\) and \(Q\) satisfying \eqref{e:Lyapunov equation}, then \(0\) is asymptotically stable.
				\end{itemize}
			\item Beyond the regime of linear systems, we have the \texttt{SOSTOOLS} library \cite{ref:SOSTOOLS, ref:TanPac-08, ref:AndPap-15, ref:JonPee-23, ref:AngMilPap-13, ref:MenWanYanXieGuo-20, ref:HanCheLuk-14, ref:KunAng-15, ref:ZhaSonWanXue-23}, developed over two decades with myriad applications, that caters to \emph{polynomial vector fields} \(\vecfld\) and works within the ambit of polynomial (sum of squares) Lyapunov functions \(\lyapfn\). The theory of \texttt{SOSTOOLS} is rooted in algebraic geometry, the constructive procedure for Lyapunov functions is algorithmic, and fast algorithms relying on semi-definite programming are available today; see \cite{ref:PraPapSelPar-05} and the references therein for a panoramic overview. See also the recent work \cite{ref:PolSze-22} for an interesting LMI based approach fine-tuned to rational polynomial vector fields, and the more classical work \cite{ref:VanVid-85} containing a procedure that leads to the construction of rational Lyapunov functions along with connections to Zubov's method. 
		\end{itemize}
		At present, and to the best of our knowledge, there is no simple and systematic tractable method for finding a Lyapunov function to assess stability of an equilibrium point of a nonlinear continuous vector field \(\vecfld\) in the absence of further algebraic structures.

		Against the preceding backdrop, our first contribution is a \emph{numerically tractable algorithmic procedure for the construction of Lyapunov functions for given equilibrium points of continuous vector fields} irrespective of their algebraic structure.
	\item \label{contrib:details} The following highlights of our algorithmic procedure lie beyond the ambit of current results:
		\begin{itemize}[label=\(\circ\), leftmargin=*]
			\item Finding a suitable Lyapunov function requires us to verify uncountably many inequality constraints as evidenced in \eqref{e:Lyapunov function}, and such an infinite family of constraints poses the chief difficulty in any algorithmic procedure (involving finite memory). \texttt{SOSTOOLS}, e.g., employs deep theorems in algebra that transform the verification of uncountably many inequalities into finitely many of them. In the absence of such algebraic structures, it is a priori unclear how to proceed with this task. One of the strengths of our algorithm becomes evident against the preceding backdrop and the fact that we \emph{do not stipulate the vector field \(\vecfld\) to be polynomial}. In particular, our algorithmic procedure applies (in the present context of finding Lyapunov functions) to every case that \texttt{SOSTOOLS} applies to in this context.
			\item Moreover, it turns out that \emph{the knowledge of the functional form (i.e., the analytical expression) of the vector field \(\vecfld\) is unnecessary} for our results to be applicable; the mere knowledge of the continuity of \(\vecfld\) coupled with the ability to evaluate \(\vecfld\) at will are sufficient. This particular feature is another key and unique strength of our approach,\footnote{Vector fields that do not admit explicit formulae are nonetheless important in control theory and frequently arise in, e.g., optimal control in the form of shooting functions associated with two-point boundary value problems, closed-loop systems for which the feedback is constructed by means of neural networks, etc.} and further discussion on this point appears in Remark \ref{r:no expression}.
		\end{itemize}
\end{enumerate}

\subsection{Technical approach and machinery}

Lyapunov functions for a given equilibrium point of a continuous vector field are not unique; apart from invariance relative to scaling by positive numbers, the properties in \eqref{e:Lyapunov function} are also tolerant to suitable nontrivial variations in \(\lyapfn\). This feature of `tolerance' provides a hint that the process of constructing the Lyapunov function may be approached via linear approximation algorithms,\footnote{The domain of \emph{linear approximation} consists of techniques involving the construction of approximants to a given function from a pre-specified vector subspace of some underlying Banach space.} which is at the root of our approach.

Our algorithmic procedure first casts the problem of finding an admissible Lyapunov function into the form of a convex semi-infinite program (SIP),\footnote{This particular step accounts for the uncountably many inequalities in \eqref{e:Lyapunov function}.} and then furnishes an admissible Lyapunov function by leveraging recent results pertaining to near-optimal solutions to convex SIPs \cite{ref:DasAraCheCha-22, ref:ParCha-23}. Notwithstanding the infinite family of constraints, the algorithmic approach espoused herein needs constant memory for a given problem that scales linearly with the dimension of the state-space; see Remark \ref{r:constant memory} for a detailed discussion, and the technical details may be found in \secref{s:problem}.

\subsection{Organization and notation}

\secref{s:problem} contains the technical formulation of our problem --- the abstract matter of finding a Lyapunov function for a given continuous vector field, and its equilibrium point is translated into a search over a finitely parametrized family of functions that must satisfy uncountably many constraints. The resulting mathematical problem turns out to be a convex semi-infinite program, and \secref{s:results} contains our main results centered around solving such programs. Numerical experiments are provided in \secref{s:numerics}, and we conclude in \secref{s:concl} with a discussion of potential directions emanating from this work.

Standard notations are employed here. The standard inner product on \(\R[\sysDim]\) is, for \(v, w\in\R[\sysDim]\), given by \(\inprod{v}{w} = \sum_{i=1}^{\sysDim} v_i w_i\), and its induced norm is the standard Euclidean norm \(\norm{v} = \sqrt{\inprod{v}{v}}\). Neighborhoods of a point in \(\R[\sysDim]\) are \emph{connected}, \emph{not necessarily open}, but are supersets of open sets containing the given point. If \(S\) is a finite set, then \(\size{S}\) denotes its cardinality.

\section{Problem formulation}
\label{s:problem}

Continuing with the notations of \secref{s:intro}, let \(\domain\subset\R[\sysDim]\) be a non-empty open set containing \(0\in\R[\sysDim]\), and let \(\vecfld:\domain\lra\R[\sysDim]\) be a continuous vector field for which \(0\) is an isolated equilibrium point. In this section, we establish a systematic, algorithmic, and tractable mechanism to search for candidate Lyapunov functions \(\lyapfn : \domain \lra\lcro{0}{+\infty}\) to assess the properties of Lyapunov stability and asymptotic stability of the origin \(0 \in \R[\sysDim]\) for the dynamical system \eqref{e:nonlinear system}.

As discussed in \secref{s:intro:indirect method}, Lyapunov's theorem asserts that the Lyapunov stability property in \ref{stab:lyapstable} of the equilibrium point \(0\in\R[\sysDim]\) is equivalent to finding a continuously differentiable function \(\domain \ni y \mapsto \lyapfn(y) \in \R\) and a neighborhood \(\nbhd\) of \(0\) satisfying\footnote{Recall that a \emph{neighborhood} for us need not be open, but it must contain a non-empty open subset.}
\begin{equation}
    \tag{\stability}
    \label{e:Lyapunov stability}
    \begin{aligned}
		& \lyapfn(0) = 0,\quad\text{and}\\
		& \begin{dcases}
			\lyapfn(y) > 0 & \text{for all }y\in\nbhd\setmin\set{0},\\
            \inprod{\pdv{\lyapfn}{x}(y)}{\vecfld(x)} \le 0 & \text{for all } y\in\nbhd.
        \end{dcases}
    \end{aligned}
\end{equation}
In a similar vein, the asymptotic stability property of \(0\in\R[\sysDim]\) in \ref{stab:asystable} is equivalent to the existence of positive definite functions \(\lyapfn, \stabilityMargin : \domain \ra \R\) and a neighborhood \(\nbhd\) of \(0\) satisfying
\begin{equation}
    \tag{\asymStability}
	\label{e:asymptotic stability}
    \begin{aligned}
        & \lyapfn(0) = 0 = \stabilityMargin(0), \quad \text{and}\\
		& \begin{dcases}
			\lyapfn(y) > 0 & \text{for all }y\in\nbhd\setmin\set{0},\\
            \stabilityMargin(y) > 0 & \text{for all } y \in \nbhd \setmin \set{0},\\
            \inprod{\pdv{\lyapfn}{x}(y)}{\vecfld(y)} + \stabilityMargin(y) \le 0 & \text{for all } y \in \nbhd.
        \end{dcases}
    \end{aligned}
\end{equation}
In what follows, the neighborhood \(\nbhd\) will always be a \emph{compact} subset of \(\domain\).

Recall that a function \(\rho:\lcro{0}{+\infty}\lra\lcro{0}{+\infty}\) is of \embf{class \(\classK\)} if \(\rho\) is continuous, strictly increasing, and \(\rho(0) = 0\).
The requirements of positive definiteness of a function \(\lyapfn\), namely \(\lyapfn(0) = 0\) and \(\lyapfn(y) > 0\) for all \(y \in \domain\setmin \set{0}\), can be equivalently \cite[\S\S41, 42]{ref:Hah-67} captured by means of the condition that there exists a function \(\lowerBound\in\classK\) satisfying \(\lyapfn(y) \ge \lowerBound(\norm{y})\) for all \(y\in\domain\). It is also a standard practice to stipulate the property of decrescence of Lyapunov functions --- that there exists a function \(\upperBound\in\classK\) satisfying \(\lyapfn(y) \le \upperBound(\norm{y})\) for all \(y\in\domain\). For instance, the two properties of positive definiteness and decrescence of a Lyapunov function \(\lyapfn\) are encapsulated by
\begin{equation}
    \label{e:pdf equiv}
    \begin{aligned}
		& \text{there exist class \(\classK\) functions \(\lowerBound, \upperBound\) such that}\\
        & \lowerBound(\norm{y}) \le \lyapfn(y) \le \upperBound(\norm{y}) \quad \text{for all } y \in \domain.
    \end{aligned}
\end{equation}

\subsection*{Candidate Lyapunov triplets}

Our search for suitable Lyapunov functions for the equilibrium point \(0\in\domain\) of \(\vecfld\) will be restricted to a suitably large space of functions described by means of the following data; we call this data a \embf{candidate Lyapunov triplet}:
\begin{enumerate}[label=\textup{(L\arabic*)}, align=left, widest=B, leftmargin=*]
	\item \label{d:nbhd} A compact neighborhood \(\nbhd \subset \domain\) of the origin \(0\in\R[\sysDim]\) that defines the region on which the Lyapunov criteria \eqref{e:Lyapunov function} are verified.
	\item \label{d:pdf} A triple \((\lowerBound, \upperBound, \marginBound)\) of class \(\classK\) functions such that \((\lowerBound, \upperBound)\) characterize positive definiteness and decrescence of the candidate Lyapunov function \(\lyapfn\) as in \eqref{e:pdf equiv}, and \(\marginBound\) characterizes positive definiteness of the stability margin \(\stabilityMargin\) by means of
		\[
			\marginBound(\norm{y}) \le \stabilityMargin(y)\quad\text{for all }y\in\nbhd.
		\]
	\item \label{d:dictionary} A pair \((\basisDict, \marginDict)\) of dictionaries of \emph{linearly independent} and \emph{continuously differentiable} functions \(\basisDict \Let \set{\basisFunc \suchthat i = 1, \ldots, \basisDim}\) and \(\marginDict \Let \set{\marginFunc \suchthat j = 1, \ldots, \marginDim}\) where
		\begin{align*}
			& \basisFunc : \domain \ra \R,\quad \marginFunc: \domain \ra \R, \quad\text{and}\\
			& \basisFunc(0) = 0 = \marginFunc(0)\quad \text{for each }i, j.
		\end{align*}
\end{enumerate}
Our search procedure picks candidate Lyapunov functions \(\lyapfn\) and candidate margin functions \(\stabilityMargin\) from the spans of the two dictionaries:
\[
    \lyapfn \in \linspan \basisDict  \quad \text{and} \quad \stabilityMargin \in \linspan \marginDict.
\]
In summary, a list \(\bigl( \nbhd, (\lowerBound, \upperBound, \marginBound), (\basisDict, \marginDict) \bigr)\) with the properties \ref{d:nbhd}--\ref{d:dictionary} is a candidate Lyapunov triplet.

Fixing a candidate Lyapunov triplet, we proceed to the algorithm for finding a Lyapunov function for the equilibrium point \(0\). This involves checking whether a Lyapunov function \(\lyapfn\in\linspan\basisDict\) and a stability margin \(\stabilityMargin\in\linspan\marginDict\) satisfying the stability requirements \eqref{e:Lyapunov stability} and/or asymptotic stability conditions \eqref{e:asymptotic stability} exist, and the procedure is carried out in \secref{s:results}.

\section{Our algorithmic procedure and associated results}
\label{s:results}

\begin{figure}[htpb]
	\begin{tikzpicture}[mindmap, grow = -90, every node/.style=concept, concept color=red!40, level 1/.append   style={level distance=5cm, sibling angle=120}, level 2/.append style={level distance=4.5 cm, sibling angle=160}, every annotation/.style={concept color=Cyan!15, text width={}, align=left}]

		\node (n3){
			A continuous vector field \(\vecfld:\domain\lra\R[\sysDim]\) and its equilibrium point \(0\)
    	}

		child[level distance=5cm, concept color=blue!25!]{
			node (n0) {
				Pick a candidate Lyapunov triplet
			}[clockwise from=-10]
				child[concept color=orange!30!]{
					node (n1){
						solve \eqref{e:asym global max} by means of \eqref{e:asym stability optimization}
					}
				}
				child[concept color=orange!30!]{
					node (n2){
						solve \eqref{e:stability global max} by means of \eqref{e:stability optimization}
					}
				}
		};

    	\node [annotation, right] at (n3.east) {
			Given data consisting of \\
			a continuous vector field\\
			and its isolated equilibrium point
		};

		\node [annotation, below] at (n0.south) {
			\(\bigl(\nbhd, (\lowerBound, \upperBound, \marginBound), (\basisDict, \marginDict)\bigr)\)\\
			satisfying \ref{d:nbhd}--\ref{d:dictionary}
		};

    	\node [annotation, concept color=PineGreen!45, above] at (n2.north) {
			\textbf{\textsf{Lyapunov\ stability}}
		};

		\node [annotation, below] at (n2.south) {
            \(\marginDict = \set{0}\) and \(\marginBound \equiv 0\);\\
			get \(\lyapfn\) defined in \eqref{e:stability lyapfn} via \eqref{e:stability MSA}
		};

    	\node [annotation, concept color=PineGreen!45, above] at (n1.north) {
			\textbf{\textsf{asymptotic\ stability}}
		};

		\node [annotation, below] at (n1.south) {
			get a pair \((\lyapfn, \stabilityMargin)\) \\
			defined in \eqref{e:asym lyapfn} and \eqref{e:asym marginfn}\\
			via \eqref{e:asym MSA}
		};

    
    \end{tikzpicture}
	\caption{A flowchart of our algorithmic procedure for the construction of Lyapunov functions.}
	\label{fig:proc}
\end{figure}
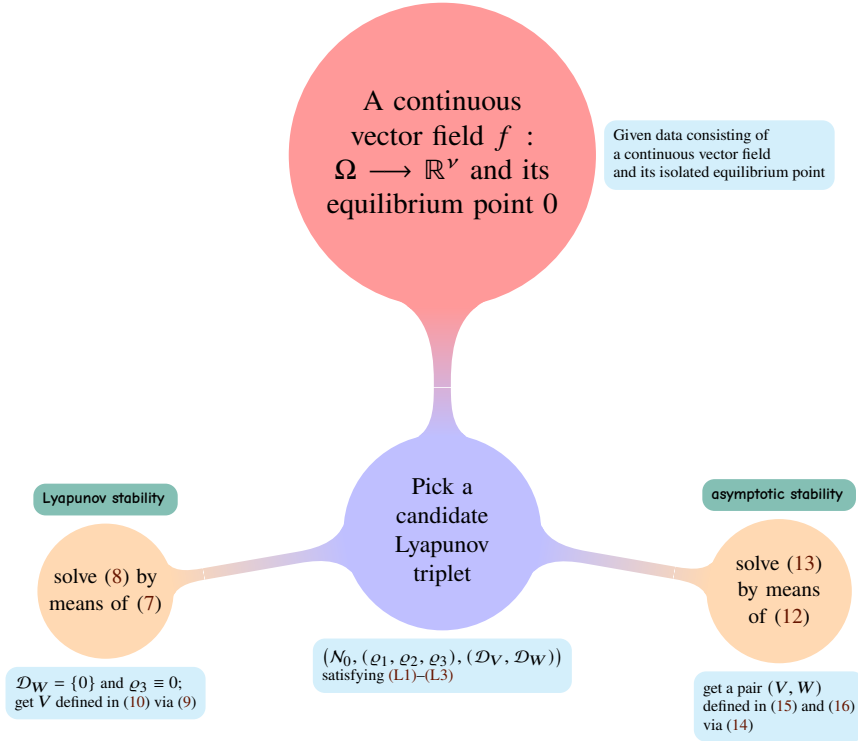

This section contains our technical results. \secref{s:results:stability} and \secref{s:results:asystability} document our results for Lyapunov stability and asymptotic stability, respectively, and \secref{s:results:discussion} contains a detailed discussion of various aspects --- theoretical and numerical --- of our results. A consolidated presentation of the proofs of our results may be found in \secref{s:results:proofs}. A schematic of the algorithmic procedure driving our results is provided in Figure \ref{fig:proc}.

\subsection{Procedure for Lyapunov stability}
\label{s:results:stability}

We begin with the problem of constructing a Lyapunov function for checking \embf{Lyapunov stability} of the (isolated) equilibrium point \(0\) of the given continuous vector field \(\vecfld\). The first step is to pick a candidate Lyapunov triplet \(\bigl( \nbhd, (\lowerBound, \upperBound, \marginBound), (\basisDict, \marginDict) \bigr)\) satisfying \ref{d:nbhd}--\ref{d:dictionary}. Since our interest is in the property of Lyapunov stability, \(\marginDict = \set{0}\) and \(\marginBound \equiv 0\).\footnote{Stricly speaking, the function \(\marginBound\) cannot be in \(\classK\) once it is set to \(0\), but we shall ignore this technicality in the interest of a cogent presentation.}

Let \(\basisDim \Let \size{\basisDict}\), and fix a strictly convex, smooth, and near-monotone function \(\objective:\R[\basisDim]\lra\lcro{0}{+\infty}\).\footnote{Recall that \emph{near-monotonicity} of $\objective$ means $\lim_{\norm{y}\to+\infty} \objective(y) = +\infty$.} For instance, the convex quadratic function \(\R[\basisDim]\ni z\mapsto \objective(z) \Let \norm{z - p}^2\) for some fixed nonzero \(p\in\R[\basisDim]\) is perfectly fine. The preceding properties on the objective \(\objective\) are designed to aid the numerical routines in our procedure.

Let \(\lyapCoeff[] \Let \pmat{\lyapCoeff[1], \ldots, \lyapCoeff[\basisDim]} \in \R[\basisDim]\) denote the coefficients of a candidate Lyapunov function in \(\linspan \basisDict\). Given the data \ref{d:nbhd}--\ref{d:dictionary}, we recast the problem of finding a feasible candidate \(\lyapfn \in \linspan \basisDict\) satisfying \eqref{e:Lyapunov stability} into the following convex semi-infinite program on the coefficients \(\lyapCoeff[]\):
\begin{equation}
    \label{e:stability csip}
    \begin{aligned}
        & \minimize_{\lyapCoeff[] \in \R[\basisDim]} && \objective(\lyapCoeff[])\\
        & \sbjto && \begin{dcases}
            \lowerBound(\norm{y}) \le \sum_{i=1}^{\basisDim} \basisFunc(y) \lyapCoeff \le \upperBound(\norm{y}) & \text{for all } y \in \nbhd,\\
            \sum_{i=1}^{\basisDim} \inprod{\pdv{\basisFunc}{x}(y)}{\vecfld(y)} \lyapCoeff \le 0 & \text{for all } y \in \nbhd.
        \end{dcases}
    \end{aligned}
\end{equation}
The problem \eqref{e:stability csip}, depending on the various objects in the selected Lyapunov triplet, may or may not be feasible. However, if it \emph{is} feasible, then a solution to \eqref{e:stability csip}, say \(\lyapCoeff[\ast]\in\R[\basisDim]\), provides the Lyapunov function 
\[
	\lyapfn(y) \Let \sum_{i=1}^{\basisDim} \basisFunc(y) \lyapCoeff[\ast, i]\quad\text{for \(y\in\nbhd\)}
\]
corresponding to the equilibrium point \(0\) that verifies its stability. If \eqref{e:stability csip} is \emph{strictly} feasible (a certain Slater's condition mentioned below holds), then we shall obtain a numerical solution to \eqref{e:stability csip} in the following fashion:

Define the map \(\relObjective : \nbhd^{\basisDim} \lra\R\) given by
\begin{equation}
    \label{e:stability optimization}
	\begin{aligned}
		& \nbhd^{\basisDim}\ni (\sample[1], \ldots, \sample[\basisDim]) \mapsto \relObjective(\sample[1], \ldots, \sample[\basisDim]) \Let \\
		& \quad\begin{aligned}
				& \inf_{\lyapCoeff[] \in \R[\basisDim]} && \objective(\lyapCoeff[])\\
		        & \sbjto && \begin{dcases}
            		\lowerBound(\norm{\sample}) \le \sum_{i=1}^{\basisDim} \basisFunc(\sample) \lyapCoeff \le \upperBound(\norm{\sample}) &  \text{for all } k = 1, \ldots, \basisDim,\\
		            \sum_{i=1}^{\basisDim} \inprod{\pdv{\basisFunc}{x}(\sample)}{\vecfld(\sample)} \lyapCoeff  \le 0 & \text{for all } k = 1, \ldots, \basisDim
       			\end{dcases}
			\end{aligned}
    \end{aligned}
\end{equation}
Observe that \eqref{e:stability optimization} contains a relaxed version of \eqref{e:stability csip} --- the number of constraints in \eqref{e:stability optimization} is \(\basisDim\) in contrast to an uncountable compact family of constraints in \eqref{e:stability csip}.

This brings us to our first main result, a proof of which is deferred to \secref{s:results:proofs}.
\begin{theorem}
	\label{t:stability}
	Let \(\domain\subset\R[\sysDim]\) be an open set containing \(0\in\R[\sysDim]\), and let a continuous vector field \(\vecfld:\domain\lra\R[\sysDim]\) be given. Suppose that \(0\) is an isolated equilibrium point of \(\vecfld\), and pick a candidate Lyapunov triplet \(\bigl( \nbhd, (\lowerBound, \upperBound), (\basisDict, \marginDict) \bigr)\) satisfying \ref{d:nbhd}--\ref{d:dictionary}. Consider the problem \eqref{e:stability csip} and suppose that the interior of its admissible set is non-empty. For \(\relObjective\) defined in \eqref{e:stability optimization}, suppose that \(\pmat{\sample[1] \opt, \ldots, \sample[\basisDim]\opt} \in \nbhd^{\basisDim}\) solves the global optimization problem
	\begin{equation}
		\label{e:stability global max}
		\sup_{(\sample[1], \ldots, \sample[\basisDim])\in\nbhd^{\basisDim}} \relObjective(\sample[1], \ldots, \sample[\basisDim]),
	\end{equation}
	and let \(\lyapCoeff[\ast]\) solve the finitely constrained convex optimization problem
    \begin{equation}
        \label{e:stability MSA}
        \begin{aligned}
            & \minimize_{\lyapCoeff[] \in \R[\basisDim]} && \objective(\lyapCoeff[])\\
            & \sbjto && \begin{dcases}
                \lowerBound \bigl(\norm{\sample \opt} \bigr) \le \sum_{i=1}^{\basisDim} \basisFunc(\sample \opt) \lyapCoeff \le \upperBound \bigl(\norm{\sample \opt} \bigr) & \text{for all } k = 1, \ldots, \basisDim,\\
                \sum_{i=1}^{\basisDim} \inprod{\pdv{\basisFunc}{x}(\sample \opt)}{\vecfld(\sample \opt)} \lyapCoeff  \le 0 & \text{for all } k = 1, \ldots, \basisDim.
            \end{dcases}
        \end{aligned}
    \end{equation}
	Then 
	\begin{equation}
		\label{e:stability lyapfn}
		\nbhd\ni z\mapsto \lyapfn(z) \Let \sum_{i=1}^{\basisDim} \basisFunc(z) \lyapCoeff[\ast, i]\in\R[]
	\end{equation}
	is a Lyapunov function for \(0\in\R[\sysDim]\) and \(0\) is Lyapunov stable.
\end{theorem}

\subsection{Procedure for asymptotic stability}
\label{s:results:asystability}

The procedure for constructing a Lyapunov function to check \embf{asymptotic stability} of the (isolated) equilibrium point \(0\) of the given continuous vector field \(\vecfld\) follows analogously as above. Pick a candidate Lyapunov triplet \(\bigl( \nbhd, (\lowerBound, \upperBound, \marginBound), (\basisDict, \marginDict) \bigr)\) satisfying \ref{d:nbhd}--\ref{d:dictionary}.

Define \(\basisDim \Let \size{\basisDict}, \marginDim \Let \size{\marginDict}\), and choose a strictly convex, smooth, and near-monotone function \(\objective:\R[\basisDim] \times \R[\marginDim] \lra \lcro{0}{+\infty}\). As before, the convex quadratic function \(\R[\basisDim] \times \R[\marginDim] \ni (u, v) \mapsto \objective(u, v) \Let \norm{u - p}^2 + \norm{v - q}^{2}\), for some fixed nonzero \(p\in\R[\basisDim], q \in \R[\marginDim]\), is a suitable choice.

Let \(\lyapCoeff[] \Let \pmat{\lyapCoeff[1], \ldots, \lyapCoeff[\basisDim]}, \marginCoeff[] \Let \pmat{\marginCoeff[1], \ldots, \marginCoeff[\marginDim]}\) denote the coefficients of the candidate functions \(\lyapfn \in \linspan \basisDict\) and \(\stabilityMargin \in \linspan \marginDict\) respectively. The feasibility problem of the requirements \eqref{e:asymptotic stability} can be equivalently phrased as solving the following convex semi-infinite program on the coefficients \(\lyapCoeff[], \marginCoeff[]\):
\begin{equation}
    \label{e:asymptotic csip}
    \begin{aligned}
		& \minimize_{(\lyapCoeff[], \marginCoeff[])} && \objective(\lyapCoeff[], \marginCoeff[])\\
        & \sbjto && \begin{dcases}
            \lowerBound(\norm{y}) \le \sum_{i=1}^{\basisDim} \basisFunc(y) \lyapCoeff \le \upperBound(\norm{y}) & \text{for all } y \in \nbhd,\\
            \marginBound(\norm{y}) \le \sum_{j=1}^{\marginDim} \marginFunc(y) \marginCoeff & \text{for all } y \in \nbhd,\\
            \sum_{i=1}^{\basisDim} \inprod{\pdv{\basisFunc}{x}(y)}{\vecfld(y)} \lyapCoeff + \sum_{j=1}^{\marginDim} \marginFunc(y) \marginCoeff \le 0 \quad & \text{for all } y \in \nbhd,\\
			(\lyapCoeff[], \marginCoeff[]) \in \R[\basisDim]\times\R[\marginDim].
        \end{dcases}
    \end{aligned}
\end{equation}

Feasibility of the problem \eqref{e:asymptotic csip} is subject to the choice of the Lyapunov triplet (see Remark \ref{r:choice} for a discussion). However, the solution \((\lyapCoeff[\ast], \marginCoeff[\ast]) \in \R[\basisDim] \times \R[\marginDim]\) to \eqref{e:asymptotic csip}, if it exists, readily produces a Lyapunov function
\[
	\lyapfn(y) \Let \sum_{i=1}^{\basisDim} \basisFunc(y) \lyapCoeff[\ast, i] \quad\text{for \(y\in\nbhd\)}
\]
which verifies the asymptotic stability of the equilibrium point \(0\) with the stability margin given by the positive definite function
\[
    \stabilityMargin(y) \Let \sum_{j=1}^{\marginDim} \marginFunc(y) \marginCoeff \quad \text{for } y \in \nbhd.
\]
In addition, if \eqref{e:asymptotic csip} satisfies a certain Slater's condition (to be mentioned below), then the solution \((\lyapCoeff[\ast], \marginCoeff[\ast])\) maximizes the map \(\relObjective : \nbhd^{\basisDim + \marginDim} \lra\R\) defined by
\begin{equation}
    \label{e:asym stability optimization}
	\begin{aligned}
		& \nbhd^{\basisDim + \marginDim} \ni (\sample[1], \ldots, \sample[\basisDim + \marginDim]) \mapsto \relObjective(\sample[1], \ldots, \sample[\basisDim + \marginDim]) \Let \\
		& \quad\begin{aligned}
			& \inf_{(\lyapCoeff[], \marginCoeff[])} && \objective(\lyapCoeff[], \marginCoeff[])\\
		        & \sbjto && \begin{dcases}
            		\lowerBound(\norm{\sample}) \le \sum_{i=1}^{\basisDim} \basisFunc(\sample) \lyapCoeff \le \upperBound(\norm{\sample}) &\\ 
                    \marginBound(\norm{\sample}) \le \sum_{j=1}^{\marginDim} \marginFunc(\sample) \marginCoeff & \text{for all } k = 1, \ldots, \basisDim + \marginDim,\\
                    \sum_{i=1}^{\basisDim} \inprod{\pdv{\basisFunc}{x}(y)}{\vecfld(\sample)} \lyapCoeff + \sum_{j=1}^{\marginDim} \marginFunc(\sample) \marginCoeff \le 0 & \\
					(\lyapCoeff[], \marginCoeff[]) \in \R[\basisDim]\times\R[\marginDim].
       			\end{dcases}
			\end{aligned}
    \end{aligned}
\end{equation}

\begin{theorem}
	\label{t:asym stability}
	Let \(\domain\subset\R[\sysDim]\) be an open set containing \(0\in\R[\sysDim]\), and let a continuous vector field \(\vecfld:\domain\lra\R[\sysDim]\) be given. Suppose that \(0\) is an isolated equilibrium point of \(\vecfld\), and pick a candidate Lyapunov triplet \(\bigl( \nbhd, (\lowerBound, \upperBound, \marginBound), (\basisDict, \marginDict) \bigr)\) satisfying \ref{d:nbhd}--\ref{d:dictionary}. Consider the problem \eqref{e:asymptotic csip} and suppose that the interior of its admissible set is non-empty. For \(\relObjective\) defined in \eqref{e:asym stability optimization}, suppose that \(\pmat{\sample[1] \opt, \ldots, \sample[\basisDim + \marginDim]\opt} \in \nbhd^{\basisDim + \marginDim}\) solves the global optimization problem
	\begin{equation}
		\label{e:asym global max}
		\sup_{(\sample[1], \ldots, \sample[\basisDim + \marginDim])\in\nbhd^{\basisDim+\marginDim}} \relObjective(\sample[1], \ldots, \sample[\basisDim + \marginDim]),
	\end{equation}
    and let \((\lyapCoeff[\ast], \marginCoeff[\ast])\) solve the finitely constrained convex optimization problem
    \begin{equation}
        \label{e:asym MSA}
        \begin{aligned}
			& \minimize_{(\lyapCoeff[], \marginCoeff[])} && \objective(\lyapCoeff[], \marginCoeff[])\\
            & \sbjto && \begin{dcases}
            		\lowerBound(\norm{\sample \opt}) \le \sum_{i=1}^{\basisDim} \basisFunc(\sample \opt) \lyapCoeff \le \upperBound(\norm{\sample \opt}) &\\ 
                    \marginBound(\norm{\sample \opt}) \le \sum_{j=1}^{\marginDim} \marginFunc(\sample \opt) \marginCoeff & \text{for all } k = 1, \ldots, \basisDim + \marginDim,\\
                    \sum_{i=1}^{\basisDim} \inprod{\pdv{\basisFunc}{x}(y)}{\vecfld(\sample \opt)} \lyapCoeff + \sum_{j=1}^{\marginDim} \marginFunc(\sample \opt) \marginCoeff \le 0 & \\ 
					(\lyapCoeff[], \marginCoeff[])\in\R[\basisDim]\times\R[\marginDim].
            \end{dcases}
        \end{aligned}
    \end{equation}
	Then 
	\begin{equation}
		\label{e:asym lyapfn}
		\nbhd \ni z \mapsto \lyapfn(z) \Let \sum_{i=1}^{\basisDim} \basisFunc(z) \lyapCoeff[\ast, i] \in \R[]
	\end{equation}
	is a Lyapunov function for \(0 \in \R[\sysDim]\) and \(0\) is asymptotically Lyapunov stable with the stability margin characterized by
    \begin{equation}
        \label{e:asym marginfn}
		\nbhd \ni z \mapsto \stabilityMargin(z) \Let \sum_{i=1}^{\marginDim} \marginFunc(z) \marginCoeff[\ast, i] \in \R[].
    \end{equation}
\end{theorem}

\subsection{Discussion}
\label{s:results:discussion}

The sequence of remarks given below documents a number of distinctive features of the results presented above.

\begin{remark}[Steps to solve the convex semi-infinite programs]
	\label{r:steps}
	In abstract terms, a convex semi-infinite program is
	\begin{equation}
		\label{e:generic csip}
		\begin{aligned}
			& \minimize_\xi && \objective(\xi)\\
			& \sbjto && 
			\begin{cases}
				\constrfn(\xi, \theta) \le 0 \quad \text{for all }\theta,\\
				\xi\in\Xi\subset\R[\nu],\; \theta\in\Theta\subset\R[\mu],
			\end{cases}
		\end{aligned}
	\end{equation}
	with continuous objective \(\objective:\Xi\lra\R[]\) and continuous constraint \(\constrfn:\Xi\times\Theta\lra\R[]\) functions such that \(\objective(\cdot)\) is convex and \(\constrfn(\cdot, \theta)\) is convex for each \(\theta\), and the sets \(\Xi\) and \(\Theta\) are convex and compact, respectively. Both \eqref{e:stability csip} and \eqref{e:asymptotic csip} are of the form \eqref{e:generic csip}; in particular, the role of the set \(\Theta\) in \eqref{e:generic csip} is played by the compact neighborhood \(\nbhd\) in \eqref{e:stability csip} and \eqref{e:asymptotic csip}. Our procedure to solve \eqref{e:generic csip} passes through numerical solutions to
	\begin{equation}
		\label{e:maxmin}
		\maximize_{(\theta_1, \ldots, \theta_\nu)\in\Theta^\nu} \; \minimize_{\xi\in\Xi} \; \set[\Big]{ \objective(\xi) \suchthat \constrfn(\xi, \theta_i) \le 0 \;\text{for each }i = 1, \ldots, \nu }.
	\end{equation}
	\cite[Theorem 1]{ref:DasAraCheCha-22} ensures that the values of \eqref{e:generic csip} and \eqref{e:maxmin} are identical, and our choice of the strictly convex and near-monotone objective function \(\objective\) also leads to the extraction of optimizers of \eqref{e:generic csip} in view of \cite[Proposition 2]{ref:DasAraCheCha-22}. Both \eqref{e:stability global max} and \eqref{e:asym global max} are of the form \eqref{e:maxmin}. The inner minimization problem in \eqref{e:maxmin} is a standard convex optimization problem, and a large number of numerical routines are available to solve it. Solutions to \eqref{e:maxmin} require the employment of global maximization routines that do not require analytical expressions of the reward functions because such formulae are not available, in general, for the value of the inner minimization as a function of \((\theta_1, \ldots, \theta_\nu)\) (encoded in the function \(\relObjective\) defined in \eqref{e:stability optimization} and \eqref{e:asym stability optimization} above). We employed off-the-shelf simulated annealing routines in our numerical experiments documented in \secref{s:numerics} partially because this technique behaves well for high-dimensional problems, but our overall approach is not contingent on the employment of simulated annealing, nor do we advocate it in all cases; see \cite[Remarks 11 and 13]{ref:DasAraCheCha-22} in this context.
\end{remark}

\begin{remark}[Nature of the theorems]
	Theorem \ref{t:stability} and Theorem \ref{t:asym stability} assume that \eqref{e:stability csip} and \eqref{e:asymptotic csip} are, respectively, feasible. This is necessary to deploy the numerical routines to solve, in the abstract language of Remark \ref{r:steps}, the max-min problem \eqref{e:maxmin}. In practice, the success of our procedure is contingent on the selection of `good' candidate Lyapunov triplets (see Remark \ref{r:choice} for a discussion), which involves an educated guess concerning the class of functions in which a Lyapunov function may exist. The two convex semi-infinite programs \eqref{e:stability csip} and \eqref{e:asymptotic csip} become feasible (if \(0\) is Lyapunov stable and asymptotically stable, respectively,) under a judicious choice of candidate Lyapunov triplets, and then the procedures in Theorems \ref{t:stability} and \ref{t:asym stability} may be deployed to arrive at corresponding Lyapunov functions. On the one hand, finding \emph{a} Lyapunov function for a set of given data (continuous vector field and its equilibrium point) involves finding \emph{a function}, and is challenging. On the other hand, making an educated guess about a class of functions --- encoded by the candidate Lyapunov triplets --- in which a Lyapunov function may exist for the given data is a simpler task (by definition!). If this guess turns out to be valid, then our algorithmic procedure finds a Lyapunov function from the selected class.
\end{remark}

\begin{remark}[Feasibility and optimization]
	Finding Lyapunov functions is, strictly speaking, a \emph{feasibility problem}. The presence of the function \(\objective\) is, therefore, theoretically unnecessary, but we include it for the ease of numerical algorithms involved in our procedure and the extraction of the correct set of optimizers of \eqref{e:stability csip} and \eqref{e:asymptotic csip}; Remark \ref{r:steps} has a discussion about the relevance of \cite[Proposition 2]{ref:DasAraCheCha-22} in this connection. The spirit of our results, therefore, mimics those involving \texttt{SOSTOOLS}. However, while \texttt{SOSTOOLS} provides certificates of the infeasibility of the original problems \eqref{e:stability csip} and \eqref{e:asymptotic csip}, our approach may not be able to detect infeasibility because it relies on relaxed (finitely constrained) versions of these problems at its core.
\end{remark}

\begin{remark}[Choice of candidate Lyapunov triplets]
	\label{r:choice}
	A judicious selection of the various components of a candidate Lyapunov triplet plays a key role in the success of the numerical technique involved herein. The pair \((\basisDict, \marginDict)\) of dictionaries in \ref{d:dictionary} should be sufficiently rich in order for the numerical routine to succeed. One typically fixes a finite number of `basis' functions from families that are \emph{dense} (relative to, e.g., the uniform norm) in the set of continuously differentiable functions on a suitable compact neighborhood of \(0\). The sizes of these dictionaries are limited by the resources at one's disposal in view of the fact that the global optimization problem scales linearly with these sizes. Moreover, the vector field \(\vecfld\) also plays an important role in the selection of \((\basisDict, \marginDict)\); indeed, equilibrium points of certain vector fields do not admit polynomial Lyapunov functions, and naturally, our procedure with polynomial dictionaries would not lead to correct solutions. The Stone-Weierstrass theorem \cite[Chapter III, \S1]{ref:Lan-93} plays a subliminal but crucial part here and motivates the linear approximation viewpoint at the heart of our technique. The neighborhood \(\nbhd\) in \ref{d:nbhd} of \(0\) should not be too large in general; indeed, for verifying the stability of \(0\), selecting a neighborhood containing any other equilibrium point of the vector field \(\vecfld\) will trivially lead to failure of the numerical procedures.
\end{remark}

\begin{remark}[Convex semi-infinite programs]
	\label{r:constant memory}
	Convex semi-infinite programs have traditionally been attacked by means of i.i.d.\ randomized sampling of the so-called ``uncertainty set'' and solving the resulting convex finitary optimization problem in the hope of getting a reasonable facsimile of the optimal solutions. Such an approach suffers from at least two difficulties:
	\begin{itemize}[label=\(\circ\), leftmargin=*]
		\item A large number of samples from the ``uncertainty set'' may be needed to achieve reasonable accuracy even in low dimensions, each of which introduces one constraint into the finitary convex optimization problem. Consequently, the complexity of the finitary convex optimization problem gets inflated in the process (and unnecessarily so, it turns out), leading to large memory requirements.
		\item Moreover, simple examples \cite[\S1.5]{ref:MisChaBan-20}, \cite[\S4]{ref:DasAraCheCha-22} demonstrate that solutions via such i.i.d.\ sampling (employed, e.g., in the scenario approach \cite{ref:CamGar-18}) could be far away from optimality with high probability, especially in high dimensions; this situation arises due to the effect known as \emph{concentration of measures}. Since, in the context of our problem, the ``uncertainty set'' is a neighborhood of an equilibrium point under consideration, such an approach involving i.i.d.\ sampling of the constraints would lead to acute difficulties for high-dimensional vector fields.
	\end{itemize}
	In contrast, our algorithmic apparatus needs \emph{constant runtime memory} if appropriate global optimization methods (such as Markov chain Monte Carlo-based simulated annealing) are adopted, and then the size of the associated global optimization scales \emph{linearly} with the number of state dimensions (which, in turn, means that the procedure performs equally well for high-dimensional vector fields); both of these are key positive features.
\end{remark}

\begin{remark}[Absence of analytic formulae]
	\label{r:no expression}
	A remarkable feature of our approach is the technique is not contingent on the availability of an analytical expression of the vector field \(\vecfld\). Situations where such analytical expressions do not naturally become available are several, and we list three important ones:
	\begin{itemize}[label=\(\circ\), leftmargin=*]
		\item Deep neural networks have had strong connections with control theory in the past \cite{ref:Son-93, ref:SonSus-97}, and their recent proliferation in feedback control synthesis is being boosted by the availability of fast computational tools. Typically, a deep neural network (with continuous activation functions) is trained via optimization routines to suit pre-specified control objectives such as asymptotic stabilization, and the outcome of the training is a feedback map in terms of the activation functions that is difficult, if not impossible to write down due to the sheer complexity of the composition of the activation maps involved. The closed-loop system, however, is a continuous vector field, and tests for the stability of the equilibria of such systems are important in practice.
		\item Model predictive control is one of the most applicable constrained control techniques at present, and it is typically implemented by means of an optimization-based oracle. An optimization routine takes the current state and outputs a corresponding control action at each stage of time. The functional form inside the oracle that provides the control action (output) for each state (input) is not available in closed form, but assessing the qualitative behavior of the associated closed-loop vector field is nonetheless important.
		\item One of the key techniques in numerical optimal control is the so-called \emph{indirect method}. It employs a set of first-order necessary conditions for optimality that is furnished, e.g., by the Pontryagin maximum principle \cite[Chapter 4]{ref:Lib-12}, and effectively transfers the task of finding an extremal pair of state and co-state trajectories to finding a root of a certain multi-dimensional continuous map \(F:\R[\sysDim]\lra\R[\sysDim]\) known as the \emph{shooting function}. An essential feature of this shooting function \(F\) is that (apart from relatively trivial cases) it can be regarded as a vector field on \(\R[\sysDim]\) (as demonstrated in \cite{ref:KumSriChaNag-22}, for instance,) and it is possible to evaluate \(F\) at points picked at will by means of numerical methods (typically involving integration), but no explicit functional formula of \(F\) can be given. This is an important example of situations in which the vector field does not admit an explicit formula.
	\end{itemize}
\end{remark}

\subsection{Proofs of Theorem \ref{t:stability} and Theorem \ref{t:asym stability}}
\label{s:results:proofs}

\begin{proof}[Proof of Theorem \ref{t:stability}]
    We begin by showing that the semi-infinite program \eqref{e:stability csip} satisfies the hypotheses of \cite[Theorem 1]{ref:DasAraCheCha-22}.
    \begin{enumerate}[label=(\roman*), align=left, widest=B, leftmargin=*]
		\item The neighborhood \(\nbhd\) plays the role of the constraint index set and is compact (in view of the condition \ref{d:nbhd} of the Lyapunov triplet).
        \item The assumption of the feasible set having a non-empty interior satisfies Slater's condition on \eqref{e:stability csip}.
        \item By construction, the cost function \(\objective\) is strictly convex and continuous. 
        \item The constraint function in \eqref{e:stability csip} can be written as an inequality of the form
            \[
                \constraintMap(\lyapCoeff[], y) \le 0 \quad \text{for all } y \in \nbhd,
            \]
            where the map \(\constraintMap: \R[\basisDim] \times \domain \lra \R[2]\) is defined by
            \[
                \constraintMap (\lyapCoeff[], y) \Let \pmat{\lowerBound(\norm{y}) - \sum_{i=1}^{\basisDim} \basisFunc(y) \lyapCoeff \\ \sum_{i=1}^{\basisDim} \basisFunc(y) \lyapCoeff - \upperBound(\norm{y}) \\ \sum_{i=1}^{\basisDim} \inprod{\pdv{\basisFunc}{x}(\sample)}{\vecfld(\sample)} \lyapCoeff}
            \]
			Clearly, for each fixed \(y \in \nbhd\) the map \(\constraintMap\) is affine in the decision variable \(\lyapCoeff[]\) and hence convex in \(\lyapCoeff[]\). Further, since the vector field \(\vecfld\) is Lipschitz continuous and the functions \(\basisFunc \in \basisDict\) are continuously differentiable, \(\constraintMap\) is also jointly continuous in \(\lyapCoeff[]\) and \(y\).
    \end{enumerate}
    Hence, from \cite[Theorem 1]{ref:DasAraCheCha-22}, the solution (optimal value) of \eqref{e:stability csip} is obtained by maximizing the function \(\relObjective\) defined in \eqref{e:stability optimization}, that is by solving \eqref{e:stability global max}.

    In addition, choosing \(\objective\) to be strictly convex ensures that the relaxed minimization problem \eqref{e:stability MSA} has a unique solution. Thus, by \cite[Proposition 2]{ref:DasAraCheCha-22}, the minimizer \(\lyapCoeff[\ast]\) of the relaxed finitely constrained optimization problem \eqref{e:stability MSA} is also the solution of the SIP \eqref{e:stability csip}. In particular, \(\lyapCoeff[\ast]\) is a feasible point for \eqref{e:stability csip}. This implies that the function \(\lyapfn\) defined in \eqref{e:stability lyapfn} corresponding to \(\lyapCoeff[\ast]\) is a valid Lyapunov function that satisfies the Lyapunov stability criterion \eqref{e:Lyapunov stability} on the neighborhood \(\nbhd\) of \(0\). Consequently, stability of the equilibrium point \(0\) follows.
\end{proof}

\begin{proof}[Proof of Theorem \ref{t:asym stability}]
    The proof of Theorem \ref{t:asym stability} follows along the lines of the proof of Theorem \ref{t:stability} presented above. Under the assumption that the convex SIP \eqref{e:asymptotic csip} under consideration is strictly feasible, it also satisfies the hypotheses of \cite[Theorem 1]{ref:DasAraCheCha-22}.

    Moreover, by choosing the objective in \eqref{e:asymptotic csip} to be strictly convex in the decision variables \((\lyapCoeff[], \marginCoeff[])\), \cite[Proposition 2]{ref:DasAraCheCha-22} applies, and consequently, the solution \((\lyapCoeff[\ast], \marginCoeff[\ast])\) of \eqref{e:asym MSA} is feasible for the convex SIP \eqref{e:asymptotic csip}. This implies that the pair of functions \(\lyapfn\) and \(\stabilityMargin\) defined in \eqref{e:asym lyapfn} and \eqref{e:asym marginfn} respectively satisfy the asymptotic stability criterion \eqref{e:asymptotic stability} on the neighborhood \(\nbhd\), thereby guaranteeing asymptotic stability of the equilibrium point \(0\).
\end{proof}

\section{Numerical experiments}
\label{s:numerics}

This section presents a number of numerical experiments to illustrate the technique developed above. A few points concerning these experiments:
\begin{enumerate}[label=(\roman*), align=right, leftmargin=*, widest=iii]
    \item For the purpose of the numerical experiments, we choose the convex quadratic objective function \(\R[\basisDim]\ni z\mapsto \objective(z) \Let \norm{z - \mathbf{1}}^2\), where \(\mathbf{1} \in \R[\basisDim]\) is a vector of all ones
    \item In our simulations, we excluded the function \(\upperBound(\cdot)\) --- serving as an upper bound for the Lyapunov functions to be constructed --- from our consideration; since the number of elements in our dictionaries is finite, the decrescent property of any Lyapunov function follows immediately.
    \item In each experiment, we validated the output of our algorithm by numerically computing the minimum values of the two functions
		\[
			y\mapsto \constraintA(y) \Let \lyapfn(y) - \lowerBound(\norm{y}) \quad \text{and}  \quad y\mapsto \constraintB(y) \Let -\stabilityMargin(\ypos) - \inprod{\pdv{\lyapfn}{x}(y)}{\vecfld(y)}
		\]
		over the domain under consideration, and they were confirmed to be non-negative.
	\item \texttt{python 3.10.4} was employed with the support of the libraries \texttt{numpy 1.24.3}, \texttt{scipy 1.8.1}, and \texttt{matplotlib 3.7.1}.
		\begin{itemize}[label=\(\circ\), leftmargin=*]
			\item The `inner' convex optimization problem in each case was solved by employing the \texttt{scipy.optimize.minimize()} function with the \texttt{``SLSQP''} method as an input parameter and \texttt{max iterations = 10000}.
			\item The `outer' (global) maximization problem in each case was solved by employing the \texttt{scipy.optimize.dual\_annealing()} function with \texttt{max iterations = 30}, and a vector (of appropriate dimension) of all entries equal to \(0.5\) as the initial starting point.
			\item In each case, the outcome of the MSA algorithm (which produces a Lyapunov function) was validated against the selected candidate Lyapunov triplet with the aid of the function \texttt{scipy.optimize.differential\_evolution()} together with its default parameters.
		\end{itemize}
\end{enumerate}

\subsection{A nonlinear planar vector field}

Consider a nonlinear 2-d system given by
\begin{equation}
    \label{e:random}
    \begin{aligned}
        & \dot{\state_{1}} = 2\state_1\left(1-\frac{\state_1}{2}\right)-\state_1\state_{2}, \quad \\
        & \dot{\state_{2}} = 3\state_2\left(1-\frac{\state_2}{3}\right)-2\state_1\state_{2}.
    \end{aligned}
    \end{equation}
It can be found that \eqref{e:random} has three equilibrium points, out of which the system is asymptotically stable around the equilibrium points \((2,0)\) and \((0,3)\). A circular disc of radius \(0.2\) centered at the equilibria is chosen as the domain of investigation for the equilibrium points. After shifting the origin to the respective equilibrium points, the following Lyapunov triplets were utilized for both points.

\subsubsection*{Candidate Lyapunov triplet for our experiment}
Our selections were as follows:
\begin{description}
	\item[\ref{d:nbhd}]  \(\nbhd  \Let \set[\big]{ (\state_1, \state_2) \in \R[2] \suchthat \state_1^2 + \state_2^2  \leq 0.04 }\).
	\item[\ref{d:pdf}] \(\lowerBound, \marginBound \in \classK\) were picked as \(\lowerBound(\radius) = \frac{\radius^2}{6}\) and \(\marginBound\left(\radius\right) =  \frac{\radius^2}{12}\) for \(\radius \ge 0\).
	\item[\ref{d:dictionary}] The dictionaries were selected to be:
		\begin{align*}
			\basisDict & \Let \set[\big]{ x_1^{i_1} \cdot x_2^{i_2} \suchthat (i_1, i_2) \in\Nz[2], i_1 + i_2  = 2 },\\ 
			\marginDict & \Let \set[\big]{ x_1^{i_1} \cdot x_2^{i_2} \suchthat (i_1, i_2) \in\Nz[2], i_1 + i_2  = 2, 4}.
		\end{align*}
\end{description}

The results of our procedure are collected in Table \ref{table_2DField}.
\begin{table}[tbh]
    \begin{center}
    \renewcommand{\arraystretch}{1.5}
    \begin{tabular}{cl}
    \toprule
    Equilibrium Point & Lyapunov Function $\state\mapsto\lyapfn(\state)$ \\ \midrule
    (2,0) & ${0.981\state_1}^2 + 0.922\state_1\state_2+{1.083\state_2}^2 -3.924\state_1-1.844\state_2+3.924$\\
    (0,3) & $1.316\state_1^2 + 0.261\state_1\state_2+
    0.336\state_2^2- 0.783\state_1-2.016\state_2+3.024$ \\
    \bottomrule
    \end{tabular}
		\caption{\label{table_2DField}
			Lyapunov functions $x\mapsto\lyapfn(x)$ for the two equilibrium points corresponding to \eqref{e:random} in the \emph{original space coordinates}.}
    \end{center}
\end{table}

\begin{figure}[h]
\centering
\begin{subfigure}{.49\textwidth}
  \centering
	\includegraphics[scale=0.3]{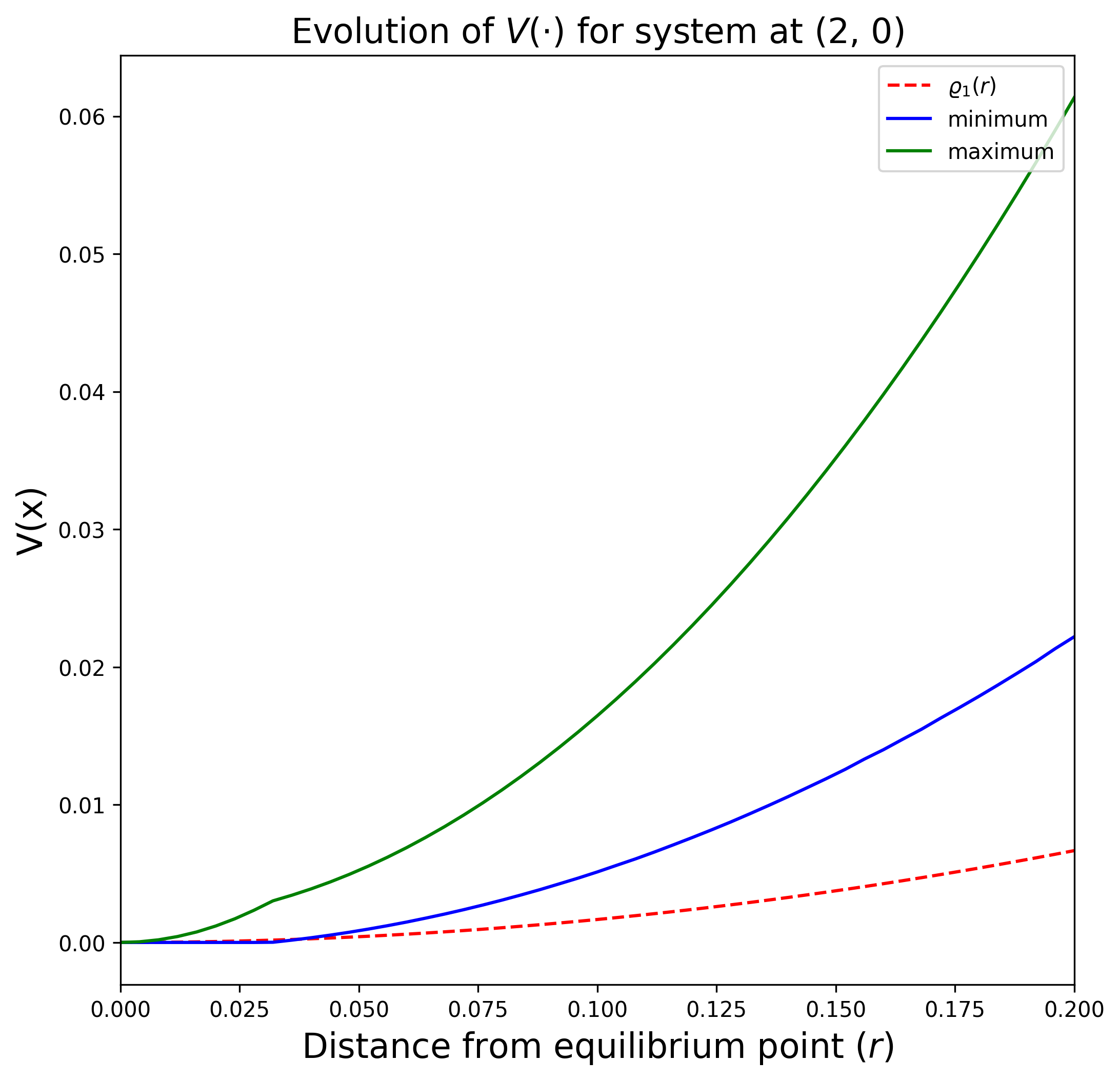}
  \caption{The minimum, on each sphere of radius \(r\), of the Lyapunov function extracted from our algorithm, lies above \(\lowerBound(\cdot)\).}
  \label{fig:2d_2-0_cand}
\end{subfigure}
\begin{subfigure}{.49\textwidth}
  \centering
  \includegraphics[scale=0.3]{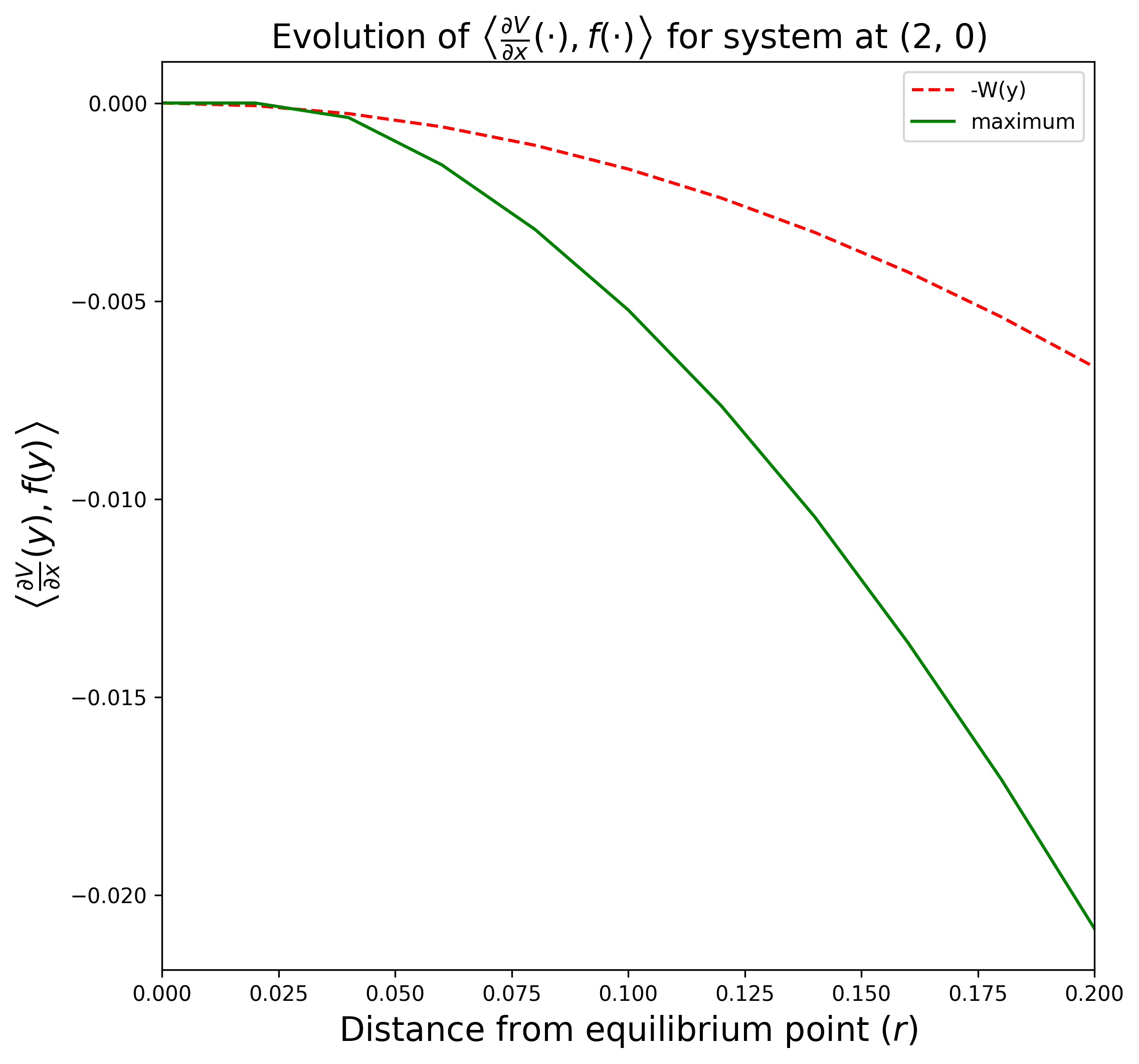}
  \caption{The maximum of \(\inprod{\pdv{\lyapfn}{x}(\cdot)}{\vecfld(\cdot)}\) lies below \(-\stabilityMargin(\cdot)\) on each sphere of radius \(r\).}
  \label{fig:2d_2-0_grad}
\end{subfigure}
	\caption{Illustration of constraint satisfaction in the case of the equilibrium point \((2, 0)\) of \eqref{e:random}.}
\label{fig:2-0}
\end{figure}
\begin{figure}[tbh]
        \centering
        \includegraphics[width=0.6\linewidth]{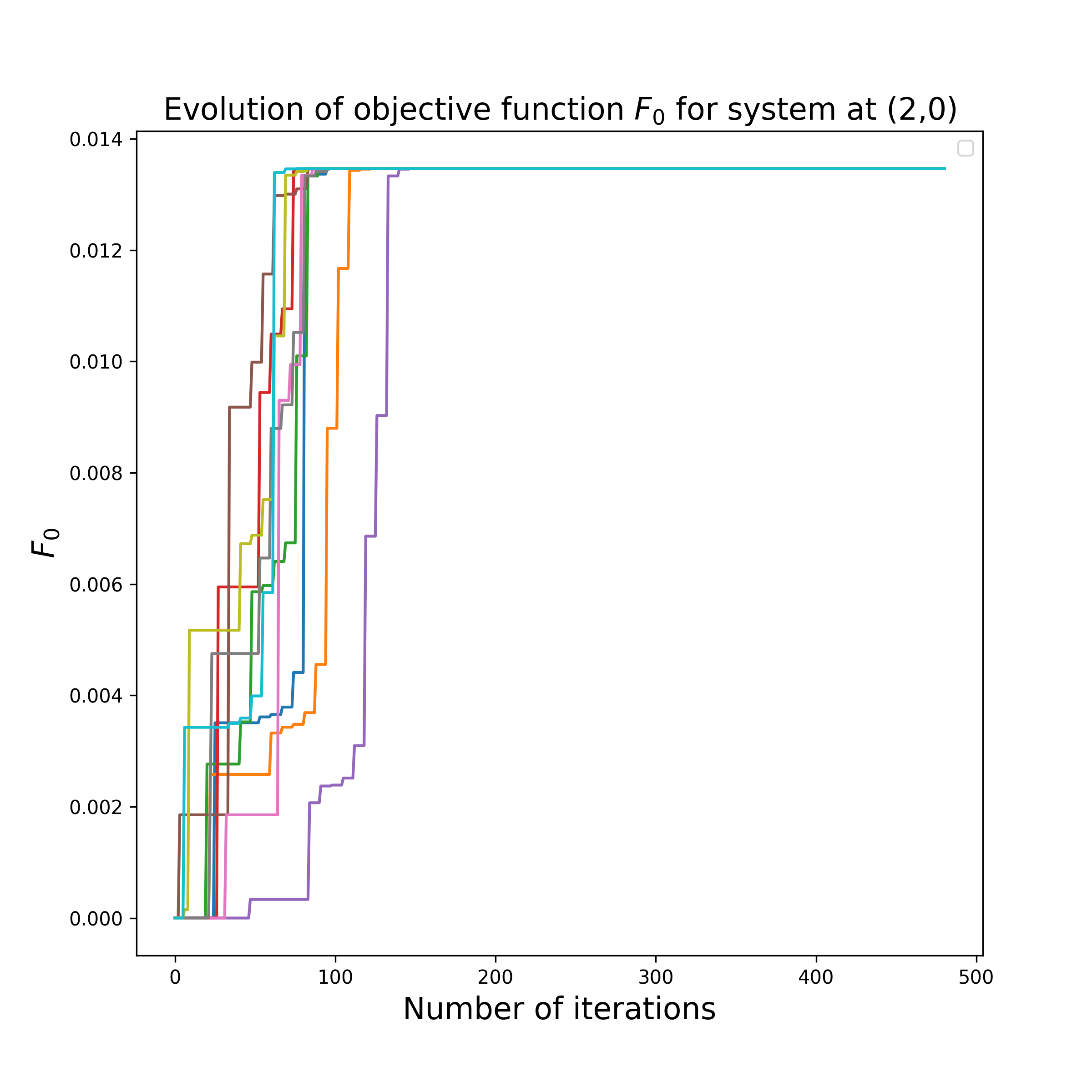}
		\caption{Illustration of the evolution of objective function \(\objective\) in the case of the equilibrium point \((2, 0)\) for \eqref{e:random} in shifted coordinates against the number of iterations over 10 executions of the simulated annealing module in our algorithm; the objective $\objective$ converges to the maximum \(0.0135\) in 100\% of the executions.}
        \label{fig: 2-0-converge}
\end{figure}

\begin{figure}[tbh]
\centering
\begin{subfigure}{.49\textwidth}
  \centering
  \includegraphics[scale=0.3]{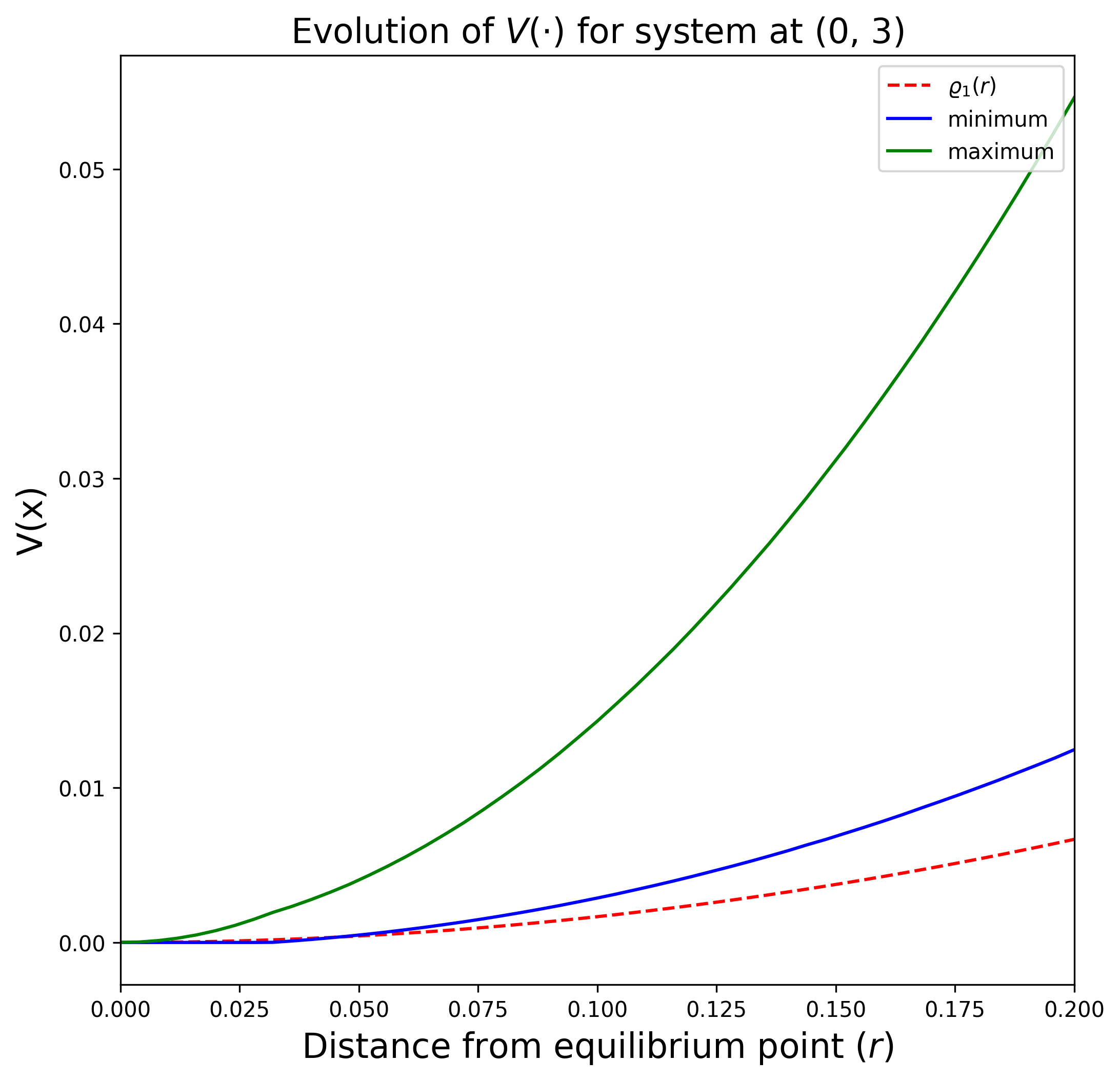}
  \caption{The minimum, on each sphere of radius \(r\), of the Lyapunov function extracted from our algorithm, lies above \(\lowerBound(\cdot)\).}
  \label{fig:0-3_cand}
\end{subfigure}
\begin{subfigure}{.49\textwidth}
  \centering
	\includegraphics[scale=0.3]{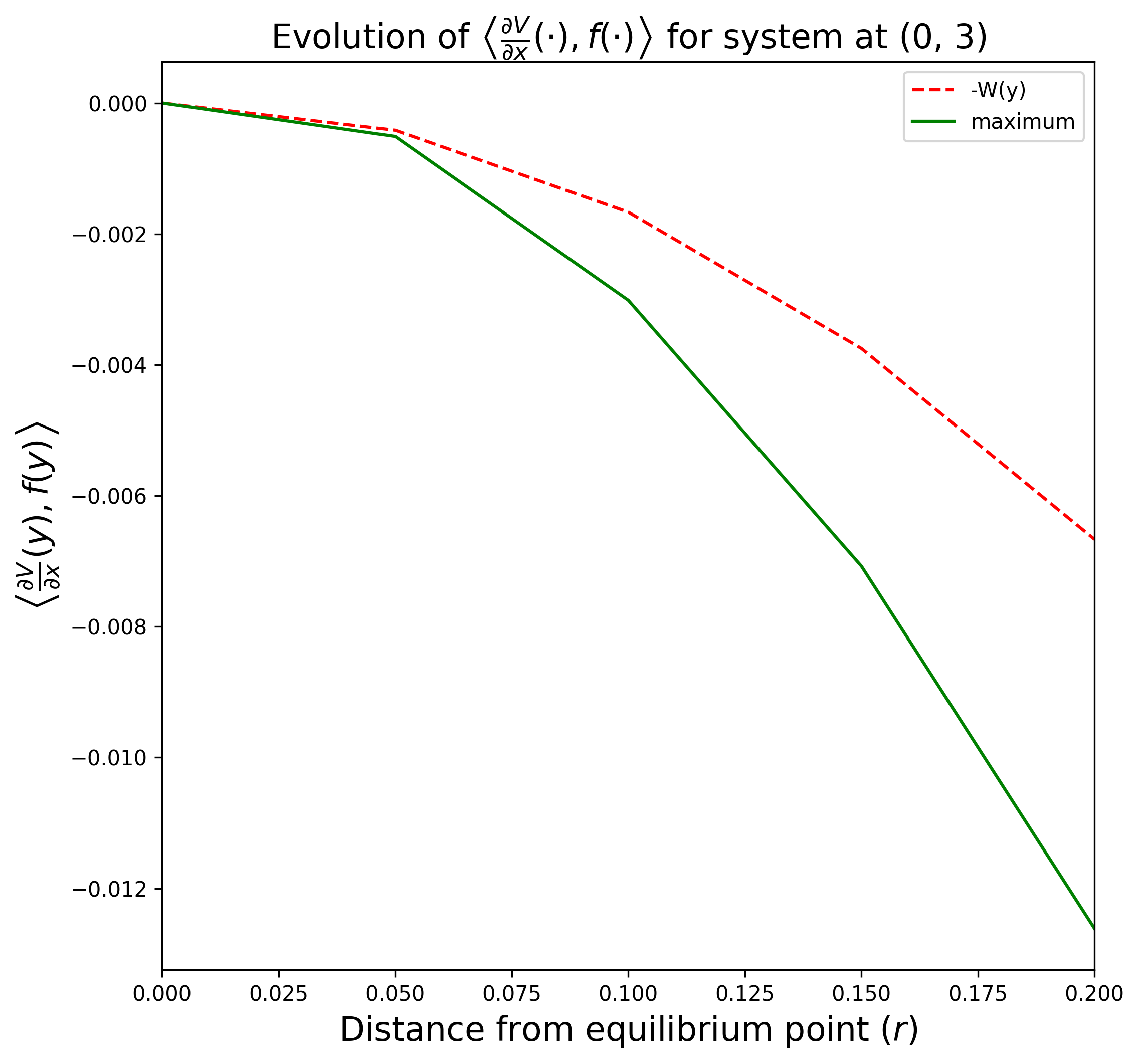}
	\caption{The maximum of \(\inprod{\pdv{\lyapfn}{x}(\cdot)}{\vecfld(\cdot)}\) lies below \(-\stabilityMargin(\cdot)\) on each sphere of radius \(r\).}
  \label{fig:0-3_grad}
\end{subfigure}
	\caption{Illustration of constraint satisfaction in the case of the equilibrium point \((0, 3)\) of \eqref{e:random}.}
\label{fig: 0-3}
\end{figure}

\begin{figure}[tbh]
        \centering
		\includegraphics[width=0.6\linewidth]{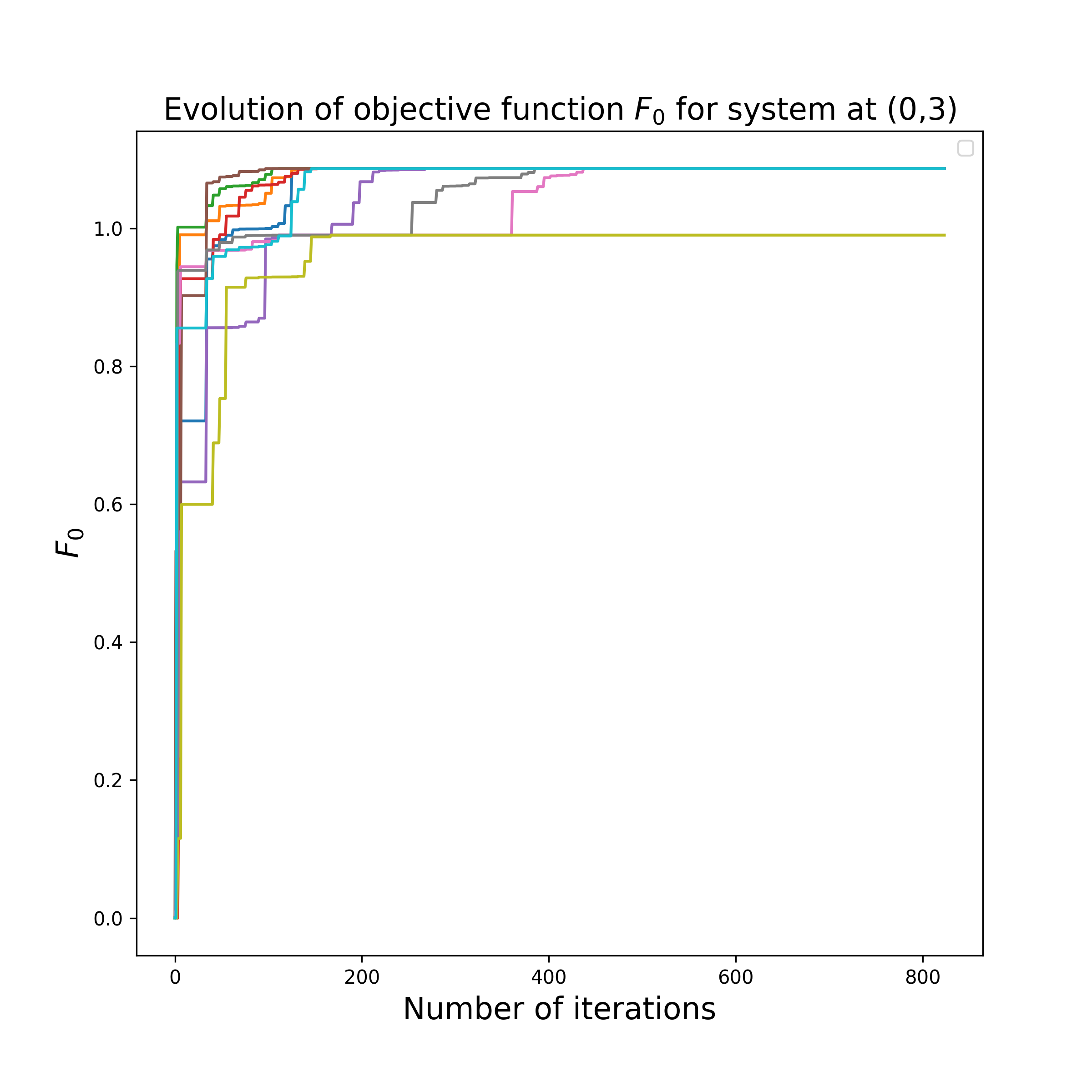}
		\caption{Illustration of the evolution of objective function \(\objective\) in the case of the equilibrium point \((0, 3)\) for \eqref{e:random} in shifted coordinates against the number of iterations over 10 executions of the simulated annealing module in our algorithm; the objective $\objective$ converges to the maximum \(1.087\) in 90\% of the executions.}
        \label{fig: 0-3-converge}
\end{figure}

\subsection{The van der Pol oscillator}

We consider a van der Pol oscillator described by
\begin{equation}
\label{e:vanderpol}
\begin{aligned}
    & \dot\state_{1} = \state_{2}, \quad \\
    & \dot\state_{2} = -\state_1 + \eps\state_2(1-\state_1^2),
\end{aligned}
\end{equation}
where \(\eps\in\R[]\) is a parameter. Observe that the stability of the above system depends on \(\eps\). To avoid an unstable limit cycle around the equilibrium point (origin), the domain is kept to be a circular disc with a radius of \(0.5\).

\subsubsection*{Candidate Lyapunov triplet for our experiment}
Our selections were as follows:
\begin{description}
	\item[\ref{d:nbhd}] \(\nbhd  \Let \set[\big]{ (\state_1, \state_2) \in \R[2] \suchthat \state_1^2 + \state_2^2  \leq 0.25 }\).
	\item[\ref{d:pdf}] \(\lowerBound, \marginBound \in \classK\) were picked as \(\lowerBound\left(\radius\right) = \frac{\radius^3}{2}\) and \(\marginBound\left(\radius\right) =  \frac{\radius^{10}}{4}\) for \(\radius \ge 0\).
	\item[\ref{d:dictionary}] The dictionaries were selected to be:
		\begin{align*}
			\basisDict & \Let \set[\big]{ x_1^{i_1} \cdot x_2^{i_2} \suchthat (i_1, i_2) \in\Nz[2],  i_1 + i_2  = 2 },\\ 
			\marginDict & \Let \set[\big]{ x_1^{i_1} \cdot x_2^{i_2} \suchthat (i_1, i_2) \in\Nz[2], i_1 + i_2  = 2, 4, \cdots, 12 }.
		\end{align*}
\end{description}
The Lyapunov function obtained from our numerical procedure was: 
\[
	\state\mapsto\lyapfn(\state) = 1.106\state_1^2 + 0.380\state_1\state_2+ 1.106\state_2^2
\]
with the value of \(\eps= -2\) for which the vector field is asymptotically stable, which happens to be the case for all \(\eps <  0\).

\begin{figure}[tbh]
\centering
\begin{subfigure}{.49\textwidth}
  \centering
  \includegraphics[scale=0.3]{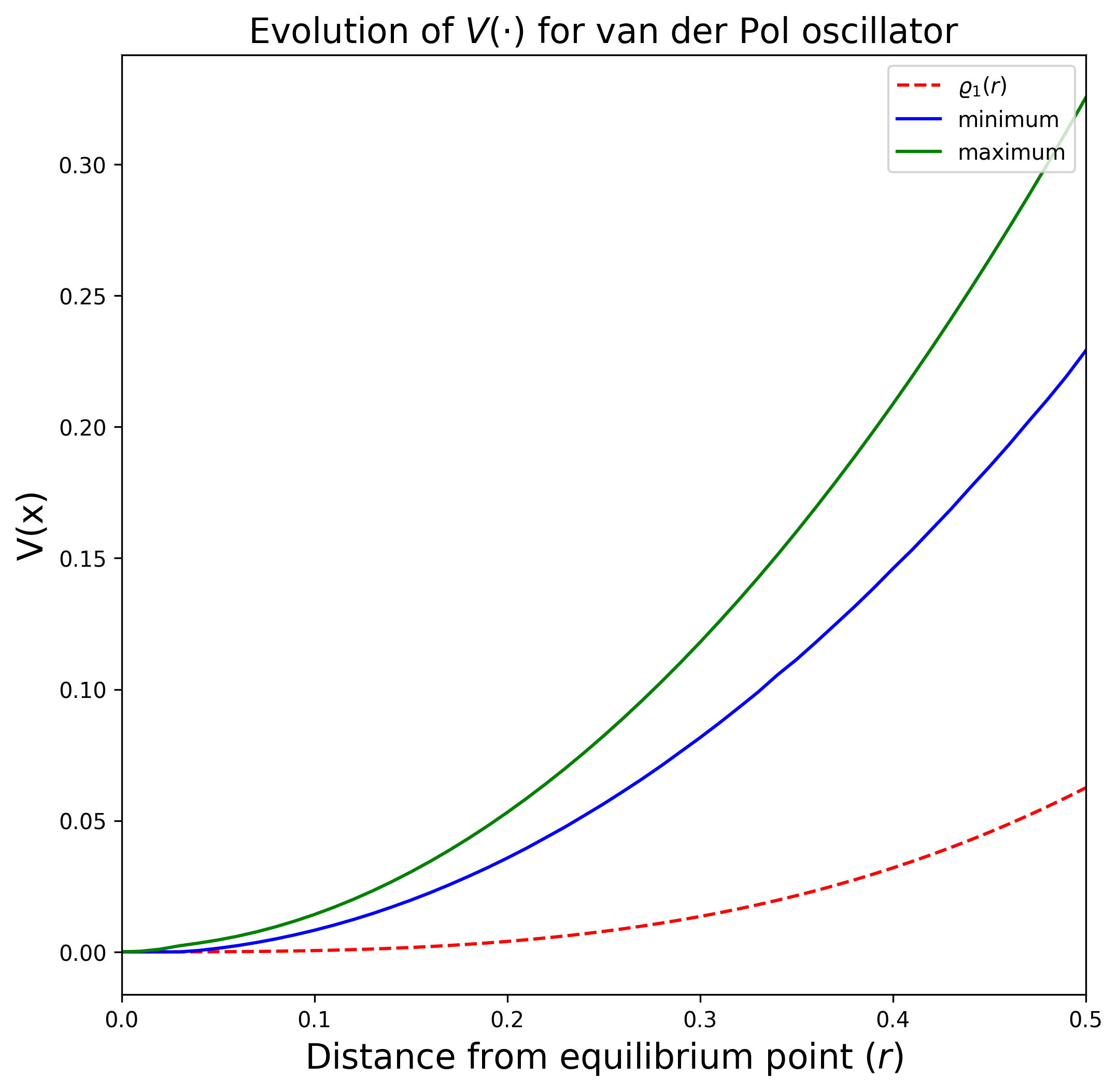}
  \caption{The minimum, on each sphere of radius \(r\), of the Lyapunov function extracted from our algorithm, lies above \(\lowerBound(\cdot)\).}
  \label{fig:vanderPol_cand}
\end{subfigure}
\begin{subfigure}{.49\textwidth}
  \centering
  \includegraphics[scale=0.3]{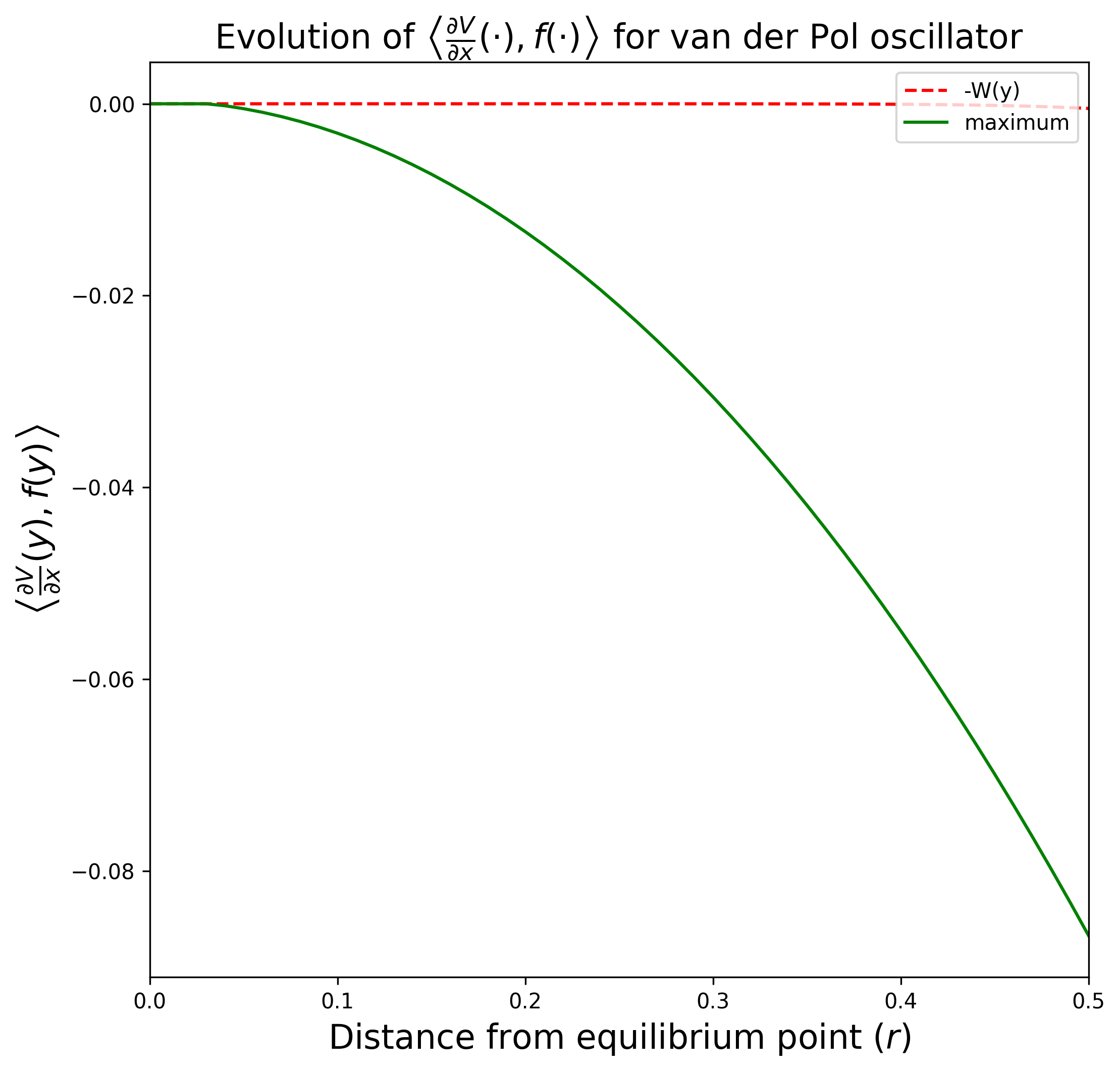}
  \caption{The maximum of \(\inprod{\pdv{\lyapfn}{x}(\cdot)}{\vecfld(\cdot)}\) lies below \(-\stabilityMargin(\cdot)\) on each sphere of radius \(r\).}
  \label{fig:vanderPol_grad}
\end{subfigure}
\caption{Illustration of constraint satisfaction in the case of \(\eps =  -2\) corresponding to \eqref{e:vanderpol}.}
\label{fig:vanderPol_lypn}
\end{figure}
\begin{figure}[tbh]
        \centering
        \includegraphics[width=0.6\linewidth]{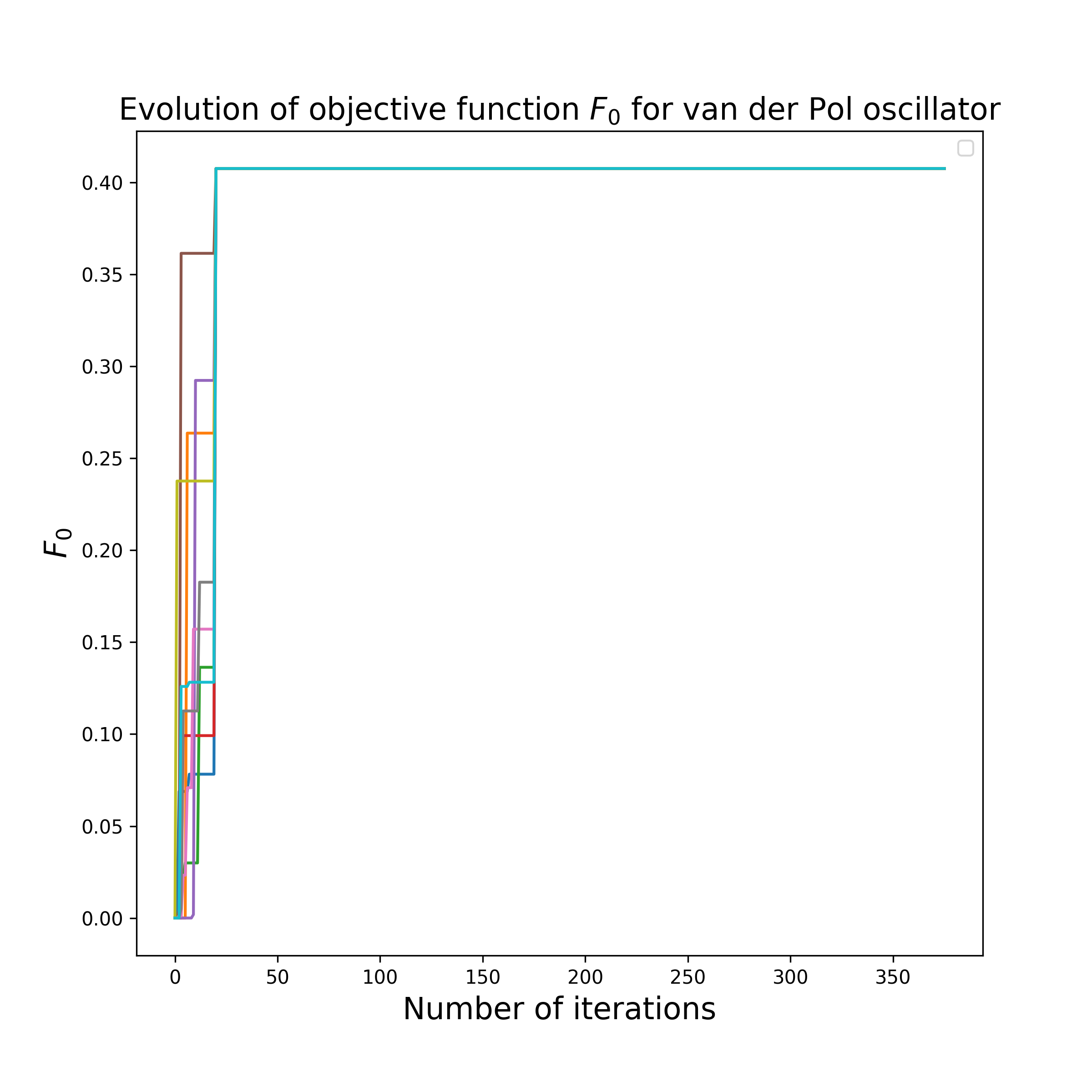}
        \caption{Illustration of the evolution of objective function \(\objective\) in the case of \(\eps =  -2\) for \eqref{e:vanderpol} against the number of iterations over 10 executions of the simulated annealing algorithm; the objective $\objective$ converges to the maximum \(0.407\) in 100\% of the executions.}
        \label{fig: vanderPol-converge}
\end{figure}

\begin{figure}[tbh]
        \centering
        \includegraphics[scale=0.6]{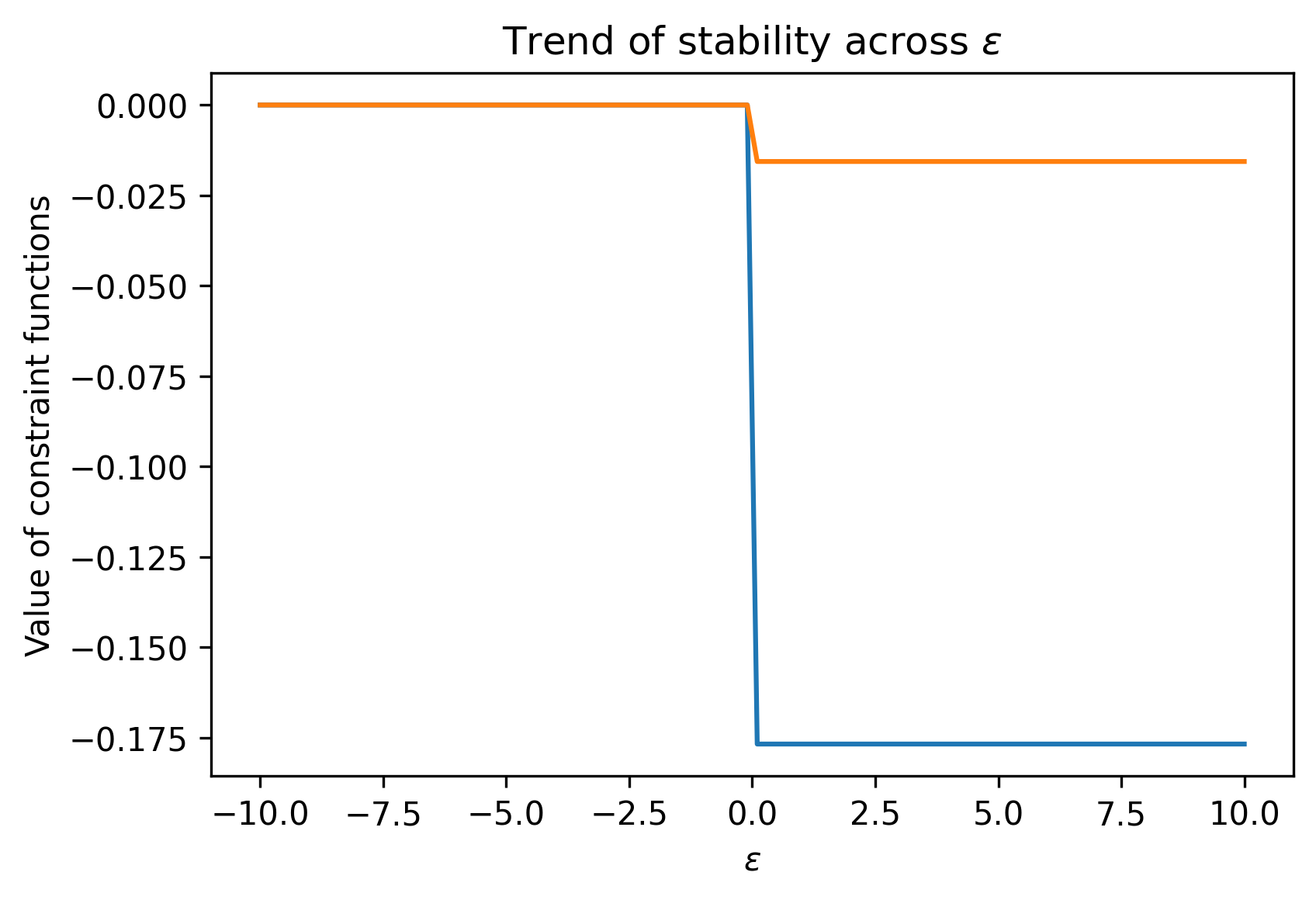}
        \caption{The variation of minimum value of both \(C_1\) and \(C_2\) in the domain indicates the system is asymptotically stable for \(\eps < 0\) and unstable otherwise.} 
        \label{fig:vary-epsilon-vanderpol}
\end{figure}

\subsection{A whirling pendulum}
Consider the whirling pendulum system lifted from \cite[Example 3]{ref:PapPra-02} and described by
\begin{equation}
    \label{e: whirl pendulum}
    \begin{aligned}
        & \dot{\state_{1}} = \state_{2},\\
        & \dot{\state_{2}} = {\dot{\theta}^2}_a\sin(\state_1)\cos(\state_1) - \frac{g}{l_a}\sin(\state_1).
    \end{aligned}
\end{equation}
The system \eqref{e: whirl pendulum} is found to be Lyapunov stable when the condition 
\begin{equation}
    \label{e:whirl pendulum condition}
	{\dot{\theta}^2}_a <\frac{g}{l_a}
\end{equation}
is met and is otherwise unstable. \cite{ref:PapPra-02} provides a Lyapunov function for the case of the above whirling pendulum system using \texttt{SOSTOOLS} after introducing two more variables to the system: $\wpstate_{1}=\sin(\state_{1})$ and $\wpstate_{2}=\cos(\state_{2})$, which transforms \eqref{e: whirl pendulum} into a 4-d polynomial vector field, and obtained a Lyapunov function of the form \(\state\mapsto a_1\state_{2}^2 + a_2\wpstate_{1}^{2} + a_3\wpstate_{2}^{2}+a_4\wpstate_{2}+a_5\). 

We now demonstrate the ability of our method to construct a Lyapunov function of a form similar to the above but without any need for  such transformations. We carried out our simulations with the numerical values of different parameters present in Table \ref{table_whirlpendulum}.
\begin{table}[tbh]
		    \begin{center}
		    \renewcommand{\arraystretch}{1.5}
			    \begin{tabular}{cc}
				    \toprule
					Parameter & Numerical Value \\
					\midrule
				    ${\dot{\theta}^2}_a$ & 1 \\
                        $\frac{g}{l_a}$ & 10\\
				    \bottomrule
			    \end{tabular}
			\end{center}
			\caption{\label{table_whirlpendulum} Parameters for the damped whirling pendulum \eqref{e: whirl pendulum}. }
\end{table}
The flexibility of our method is demonstrated in this example by the selection of both polynomial and non-polynomial functions in the dictionary for \(V(\cdot)\).

\subsubsection*{Candidate Lyapunov triplet for our experiment}

Our selections were as follows:
\begin{description}
	\item[\ref{d:nbhd}]  \(\nbhd  \Let \set[\big]{ (\state_1, \state_2) \in \R[2] \suchthat \state_1^2 + \state_2^2  \leq 1 }\).
	\item[\ref{d:pdf}] \(\lowerBound, \marginBound \in \classK\) were picked as \(\lowerBound\left(\radius\right) = \frac{\radius^2}{100}\) and \(\marginBound\left(\radius\right) =  0\) for \(\radius \ge 0\).
	\item[\ref{d:dictionary}] The dictionaries were selected to be:
		\begin{align*}
			\basisDict & \Let \set[\big]{ x_1^{i_1} \cdot x_2^{i_2} \suchthat (i_1, i_2) \in\Nz[2], i_1 + i_2  = 2 } \cup \set[\big]{ \cos\left(j_{1}x_1\right) \suchthat 0 \leq j_1 \leq 2 } ,\\ 
			\marginDict & \Let \set[\big]{ x_1^{i_1} \cdot x_2^{i_2} \suchthat (i_1, i_2) \in\Nz[2], i_1 + i_2  = 2, 4, 6 }.
		\end{align*}
\end{description}
  
\begin{figure}[tbh]
        \centering
        \includegraphics[scale = 0.6]{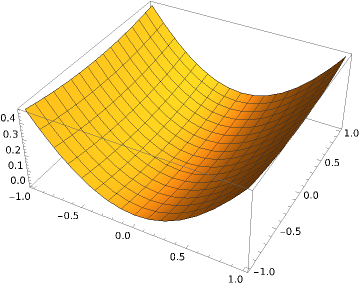}
        \caption{Lyapunov function for the whirling pendulum in \eqref{e: whirl pendulum}.} 
        \label{fig: whirling_lyapunov}
\end{figure}

The Lyapunov function for the system \eqref{e: whirl pendulum}, with parameters in Table \ref{table_whirlpendulum} and the aforementioned Lyapunov triplet, obtained from our numerical procedure was: 
\[
	\state\mapsto\lyapfn(\state) = 0.050 \state_2^2 + 0.025\cos\left(2\state_1\right)-\cos\left(\state_1\right)+0.975.
\]
Figure \ref{fig: whirling_lyapunov} depicts this function on the unit box \(\lcrc{-1}{1}^2\) (a domain containing \(\nbhd\)).

\begin{figure}[tbh]
\centering
\begin{subfigure}{.49\textwidth}
  \centering
  \includegraphics[width=\linewidth]{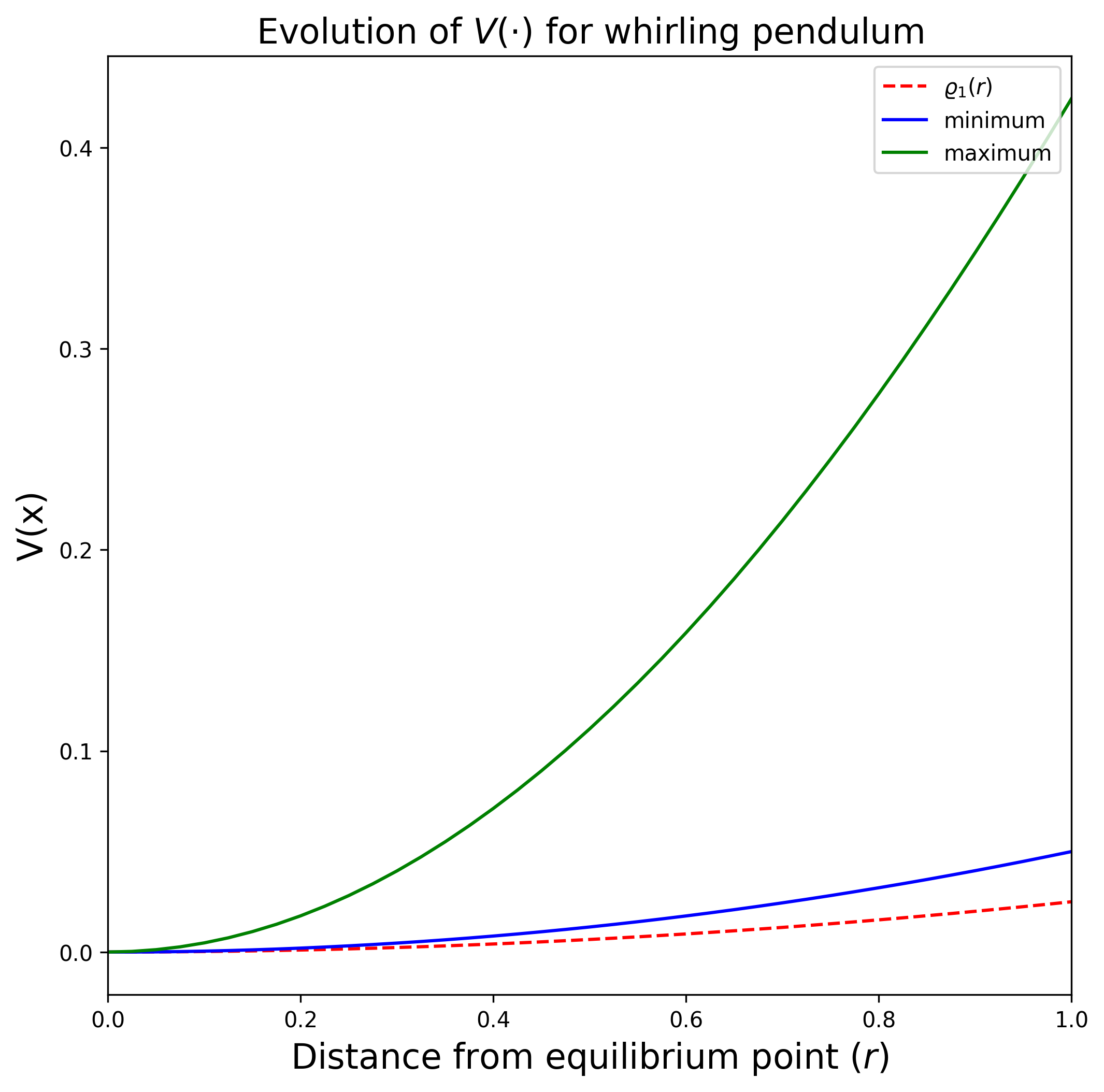}
  \caption{The minimum, on each sphere of radius \(r\), of the Lyapunov function extracted from our algorithm, lies above \(\lowerBound(\cdot)\).}
  \label{fig:whirling_cand}
\end{subfigure}
\begin{subfigure}{.49\textwidth}
  \centering
  \includegraphics[width=\linewidth]{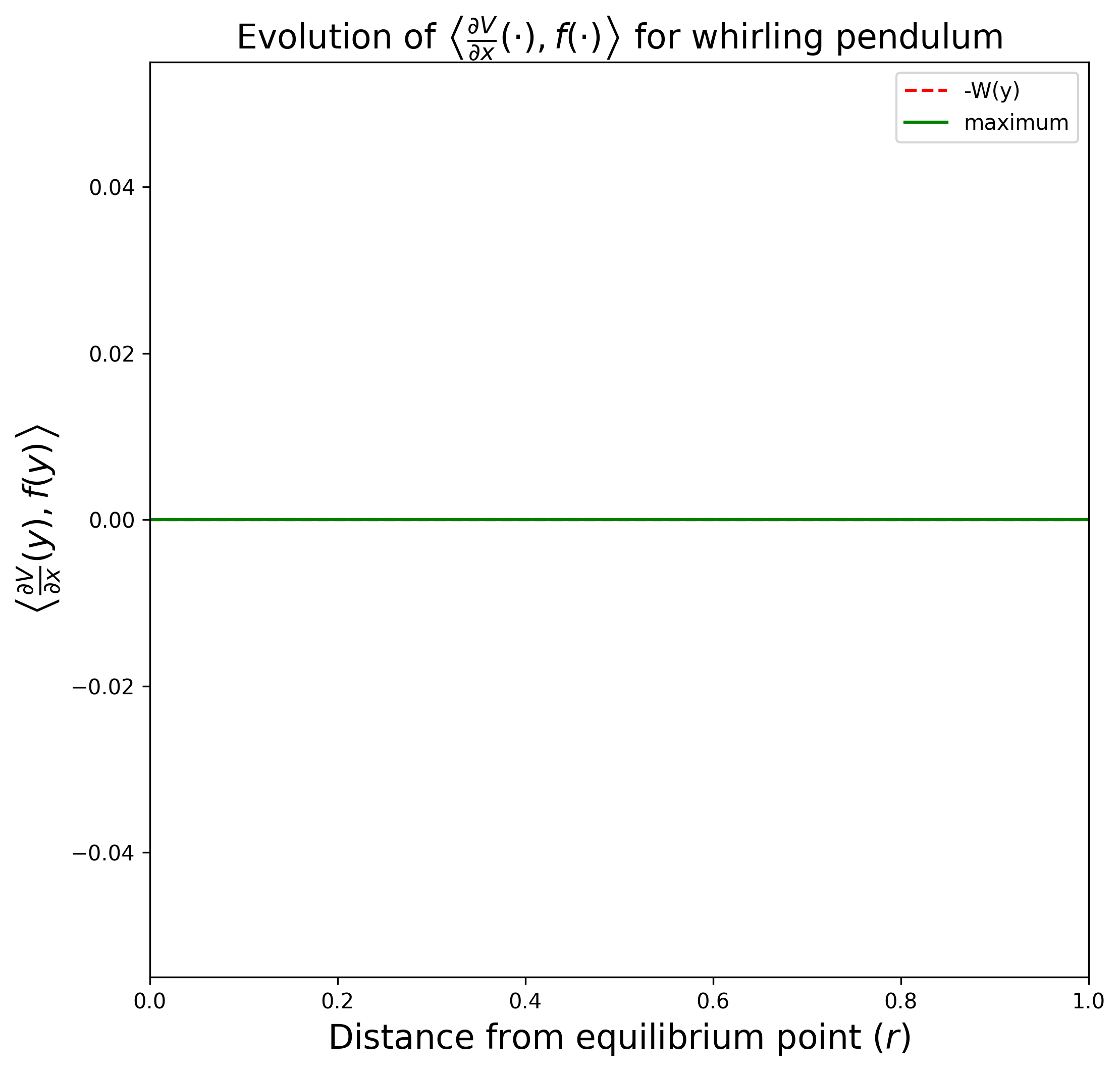}
  \caption{The maximum of \(\inprod{\pdv{\lyapfn}{x}(\cdot)}{\vecfld(\cdot)}\) lies below \(-\stabilityMargin(\cdot)\) on each sphere of radius \(r\).}
  \label{fig:whirling_grad}
\end{subfigure}
\caption{Illustration of constraint satisfaction corresponding to \eqref{e: whirl pendulum}.}
\label{fig: whirling}
\end{figure}

\begin{figure}[tbh]
        \centering
        \includegraphics[width=0.6\linewidth]{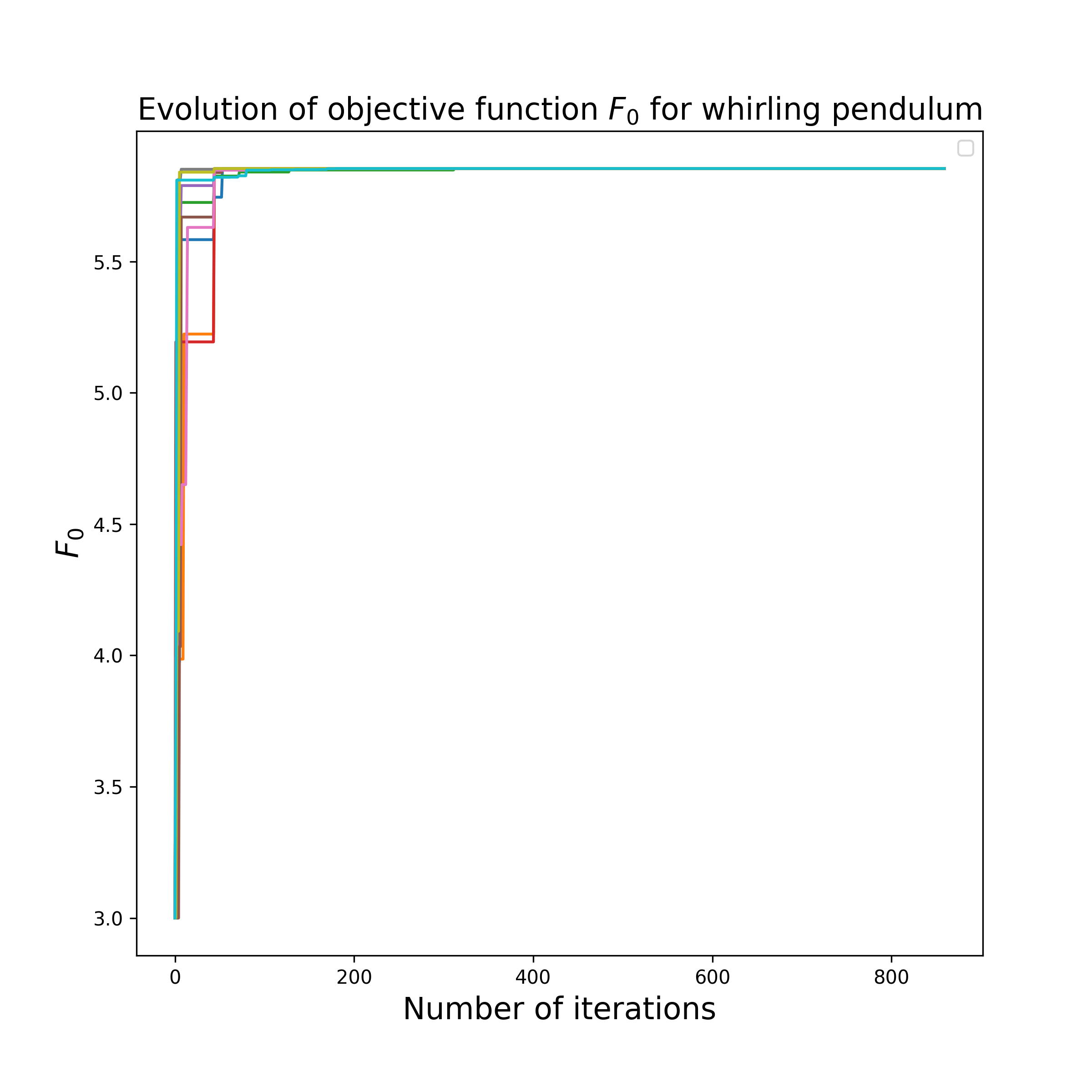}
        \caption{Illustration of the evolution of objective function \(\objective\) for \eqref{e: whirl pendulum} against the number of iterations over 10 executions of the simulated annealing algorithm; the objective $\objective$ converges to the maximum \(5.854\) in 100\% of the executions.}
        \label{fig: whirling-converge}
\end{figure}

\subsection{A 5-d hyperchaotic system with linear feedback controller}
For our next example, we introduce an illustration of a 5-d autonomous hyperchaotic system, as proposed by \cite{ref:Niu2021}. The system incorporates a linear feedback controller, effectively achieving global asymptotic stability at the origin. The system can be expressed as follows:
\begin{equation}
    \label{e:5d-system}
    \begin{aligned}
		\dot \state_{1} & = a(\state_{2} - \state_{1}) -k_1\state_1, \quad \\
		\dot \state_{2} & = (c-a)\state_1 + c\state_2 + \state_5 - \state_1\state_3 - k_2\state_2, \quad \\
		\dot \state_{3} & = -b\state_3 + \state_1\state_2 - k_3\state_3, \quad \\
		\dot \state_{4} & = m\state_5 - k_4\state_4, \quad \\
		\dot \state_{5} & = -\state_2 - h\state_4 - k_5\state_5
    \end{aligned}
\end{equation}
where \((k_{1}, k_{2}, k_{3}, k_{4},k_{5}) = (0, 30, 0, 1, 1)\). We keep \(a = 23\), \(b = 3\), \(c = 18\), \(m = 12\), and \(h = 4\) as recommended in \cite{ref:Niu2021} to obtain a global asymptotic stability at origin.

\subsubsection*{Candidate Lyapunov triplet for our experiment}
Our selections were as follows:
\begin{description}
	\item[\ref{d:nbhd}]  \(\nbhd  \Let \set[\big]{ (\state_1, \state_2, \state_3, \state_4, \state_5) \in \R[5] \suchthat \state_1, \state_2, \state_3, \state_4, \state_5 \in \lcrc{-0.5}{0.5} }\).
	\item[\ref{d:pdf}] \(\lowerBound, \marginBound \in \classK\) were picked as \(\lowerBound\left(\radius\right) = \frac{\radius^2}{100}\), and \(\marginBound\left(\radius\right) =  \frac{\radius^4}{20000}\) for \(\radius \ge 0\).
	\item[\ref{d:dictionary}] The dictionaries were selected to be:
		\begin{align*}
			\basisDict & \Let \set[\big]{ x_1^{i_1} \cdot x_2^{i_2} \cdots x_5^{i_5} \suchthat (i_1, i_2, i_3, i_4, i_5) \in\Nz[5], i_1 + i_2 + i_3 + i_4 + i_5 = 2 },\\ 
			\marginDict & \Let \set[\big]{ x_1^{i_1} \cdot x_2^{i_2} \cdots x_5^{i_5} \suchthat (i_1, i_2, i_3, i_4, i_5) \in\Nz[5], i_1 + i_2 + i_3 + i_4 + i_5  = 2, 4, 6 }.
		\end{align*}
\end{description}
Our numerical procedure led to the Lyapunov function
\begin{align*}
	\state\mapsto \lyapfn(\state) & = 1.229\state_1^2 + 0.982\state_1\state_2 + 0.891\state_1\state_3 + 0.632\state_1\state_4 + 0.236\state_1\state_5\\
	& \qquad + 0.996\state_2^2+ 1.026\state_2\state_3 + 0.663\state_2\state_4 + 0.928\state_2\state_5 + 1.116\state_3^2\\
	& \qquad + 0.406\state_3\state_4 + 1.054\state_3\state_5 + 0.523\state_4^2 + 0.158\state_4\state_5 + 1.595\state_5^2.
\end{align*}

\begin{figure}[tbh]
\centering
\begin{subfigure}{.49\textwidth}
  \centering
  \includegraphics[width=\linewidth]{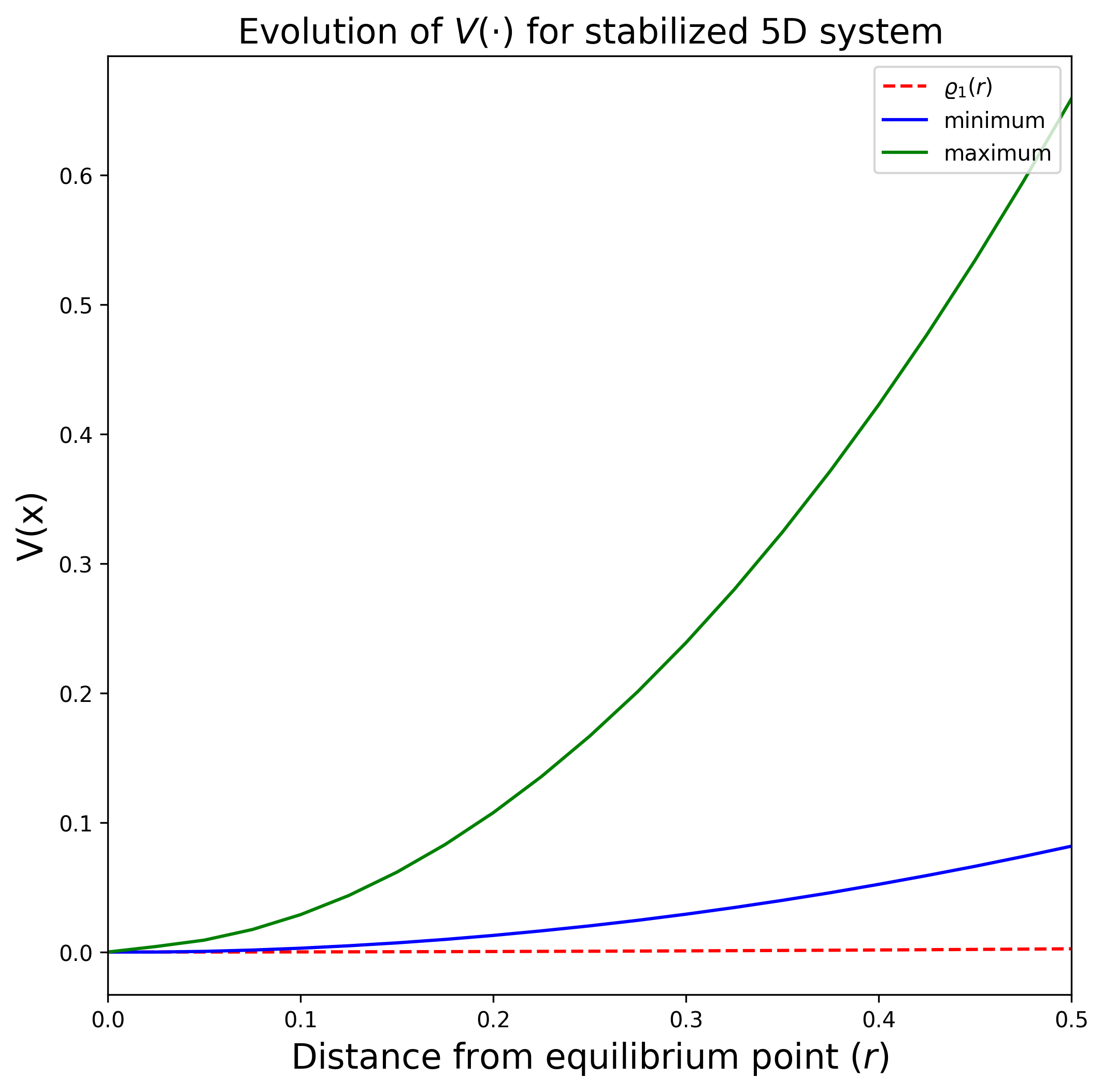}
  \caption{The minimum, on each sphere of radius \(r\), of the Lyapunov function extracted from our algorithm, lies above \(\lowerBound(\cdot)\).}
  \label{fig:5d_cand}
\end{subfigure}
\begin{subfigure}{.49\textwidth}
  \centering
  \includegraphics[width=\linewidth]{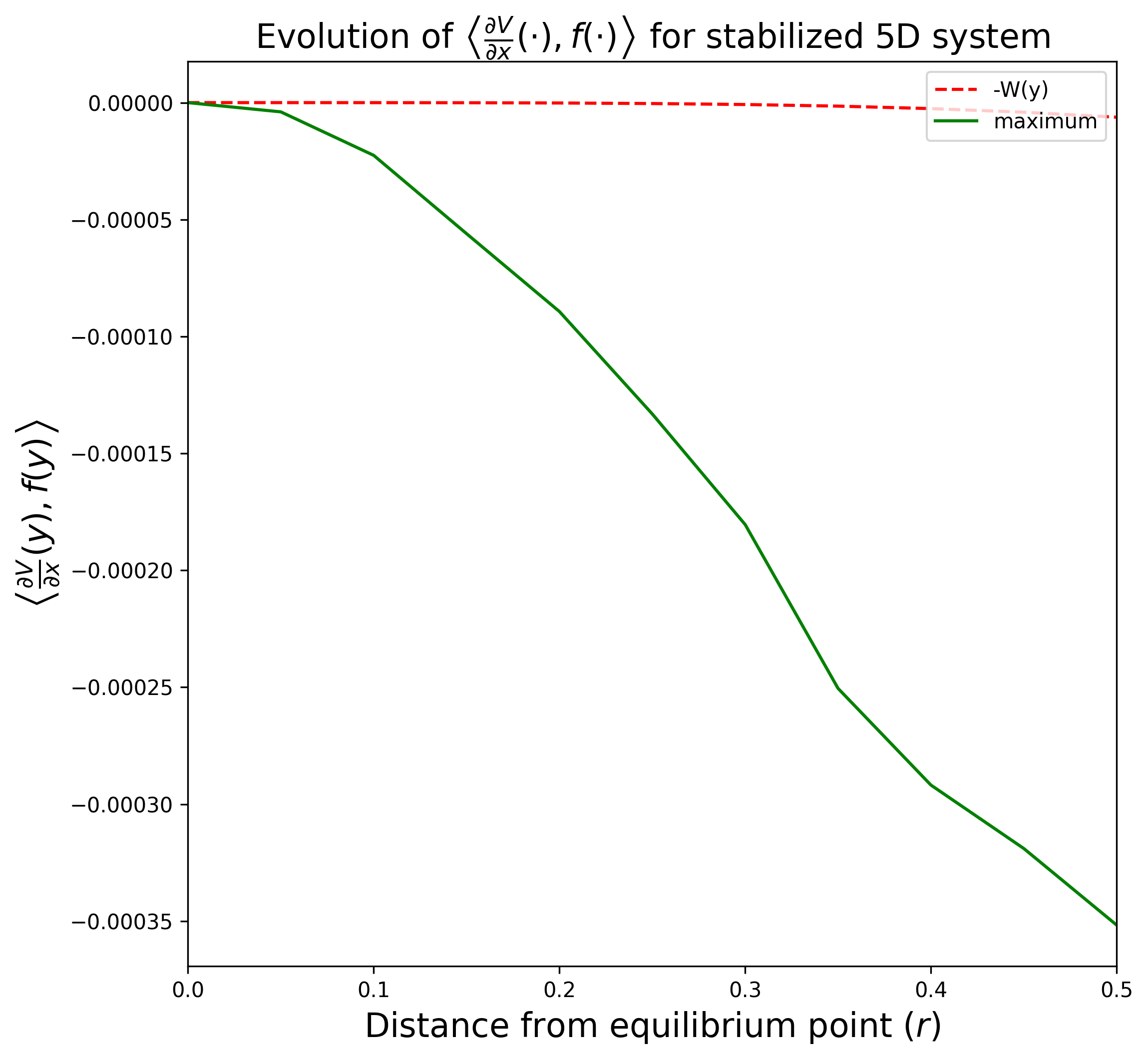}
  \caption{The maximum of \(\inprod{\pdv{\lyapfn}{x}(\cdot)}{\vecfld(\cdot)}\) lies below \(-\stabilityMargin(\cdot)\) on each sphere of radius \(r\).}
  \label{fig:5d_grad}
\end{subfigure}
\caption{Illustration of constraint satisfaction corresponding to \eqref{e:5d-system}.}
\label{fig: 5d}
\end{figure}

\begin{figure}[tbh]
        \centering
        \includegraphics[width=0.6\linewidth]{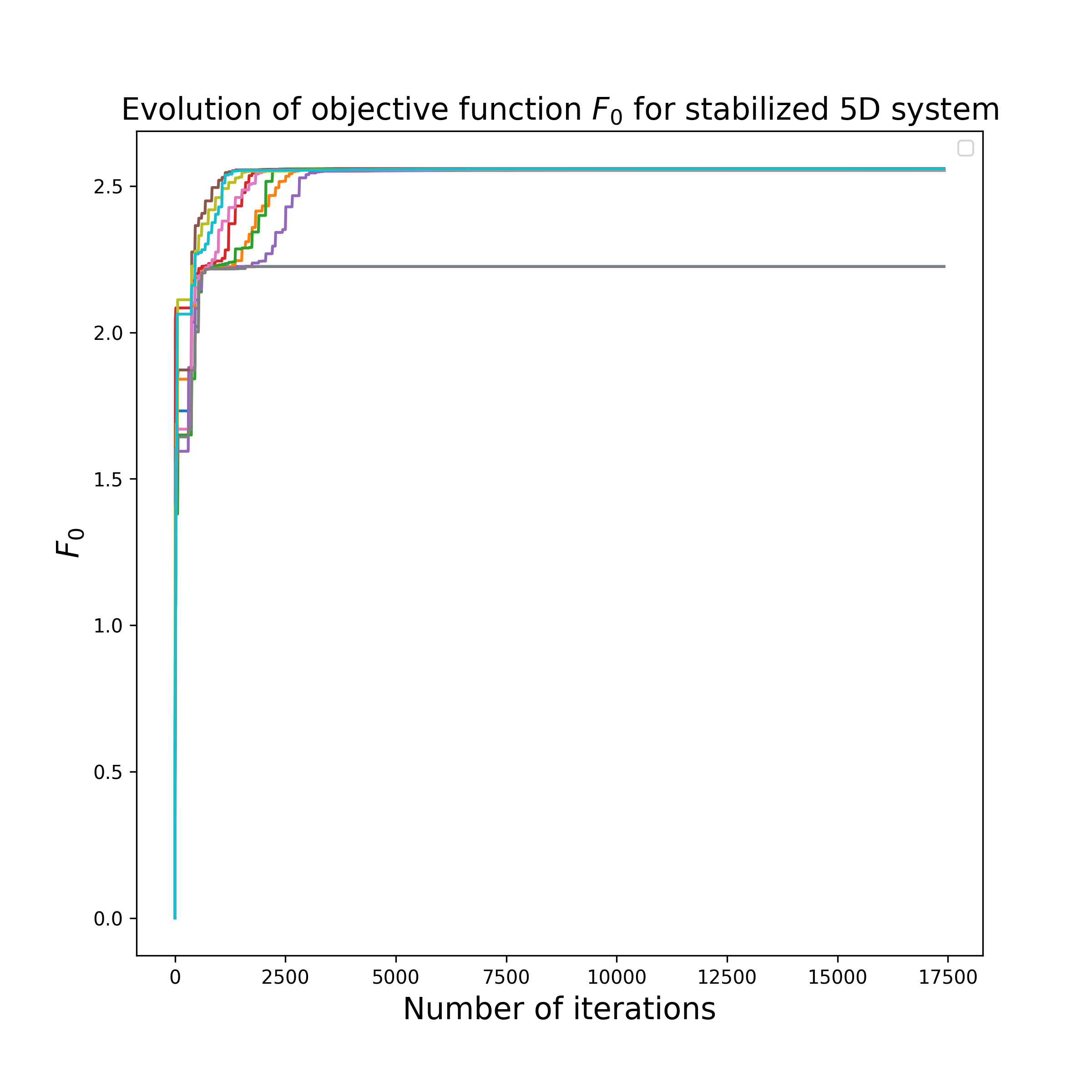}
		\caption{Illustration of the evolution of objective function \(\objective\) for \eqref{e:5d-system} against the number of iterations over 10 executions of the simulated annealing algorithm; the objective $\objective$ converges to the maximum \(2.560\) in 80\% of the executions.}
        \label{fig: 5D-converge}
\end{figure}

\subsection{Transient stability of classical power system models}

For our next example, we pick a four-dimensional system from \cite{ref:AngMilPap-13} given by the equation (transformed to have the origin as its equilibrium point):
\begin{equation}
  \label{eq:power_model}
  \begin{aligned}
	  \dot \state_{1} & = \state_{2}, \quad \\
	  \dot \state_{2} & = 0.0200\cos(\state_1)\cos(\state_3) - 0.0200\cos(\state_1) - 0.9998\sin(\state_1) - 0.4000\state_2  \\
	  & \qquad + 0.4996\cos(\state_1)\sin(\state_3) - 0.4996\cos(\state_3)\sin(\state_1) + 0.0200\sin(\state_1)\sin(\state_3), \\
	  \dot \state_{3} & = \state_{4}, \\
	  \dot \state_{4} & = 0.4996\cos(\state_3)\sin(\state_1) - 0.0299\cos(\state_3) - 0.4991\sin(\state_3) \\
    & \qquad -0.0200\cos(\state_1)\cos(\state_3) - 
    0.4996\cos(\state_1)\sin(\state_3) - 0.5000\state_4 \\
    & \qquad - 0.0200\sin(\state_1)\sin(\state_3) + 0.0500.
    \end{aligned}
\end{equation}

\subsubsection*{Candidate Lyapunov triplet for our experiment}
    Our selections were as follows:
\begin{description}
	\item[\ref{d:nbhd}]  \(\nbhd  \Let \set[\big]{ (\state_1, \state_2, \state_3, \state_4) \in \R[4] \suchthat \state_1, \state_2, \state_3, \state_4 \in \lcrc{-0.2}{0.2} }\).
	\item[\ref{d:pdf}] \(\lowerBound \in \classK\): \(\lowerBound\left(\radius\right) = \frac{\radius^2}{16}\) and \(\marginBound\left(\radius\right) =  \frac{\radius^2}{200}\) for \(\radius \ge 0\).
	\item[\ref{d:dictionary}] The dictionaries were selected to be:
		\begin{align*}
			\basisDict & \Let \set[\big]{ x_1^{i_1} \cdot x_2^{i_2} \cdot x_3^{i_3} \cdot x_4^{i_4} \suchthat (i_1, i_2, i_3, i_4) \in\Nz[4],  i_1 + i_2 + i_3 + i_4 = 2 },\\ 
			\marginDict & \Let \set[\big]{ x_1^{i_1} \cdot x_2^{i_2} \cdot x_3^{i_3} \cdot x_4^{i_4} \suchthat (i_1, i_2, i_3, i_4) \in\Nz[4], i_1 + i_2 + i_3 + i_4  = 2, 4}.
		\end{align*}
\end{description} 
This example demonstrates how our algorithm effectively handles higher order systems, enabling the identification of an appropriate Lyapunov function with predefined lower bounds. Our numerical procedure led to the Lyapunov function 
\begin{align*}
	\state\mapsto\lyapfn(\state) & = 1.339\state_1^2 + 0.591\state_1\state_2 + 0.635\state_1\state_3 + 0.622\state_1\state_4 + 0.989\state_2^2\\
	& \qquad + 0.651\state_2\state_3 + 1.063\state_2\state_4 + 1.064\state_3^2 + 0.970\state_3\state_4 + 1.297\state_4^2.
\end{align*}

\begin{figure}[tbh]
\centering
\begin{subfigure}{.49\textwidth}
  \centering
  \includegraphics[width=\linewidth]{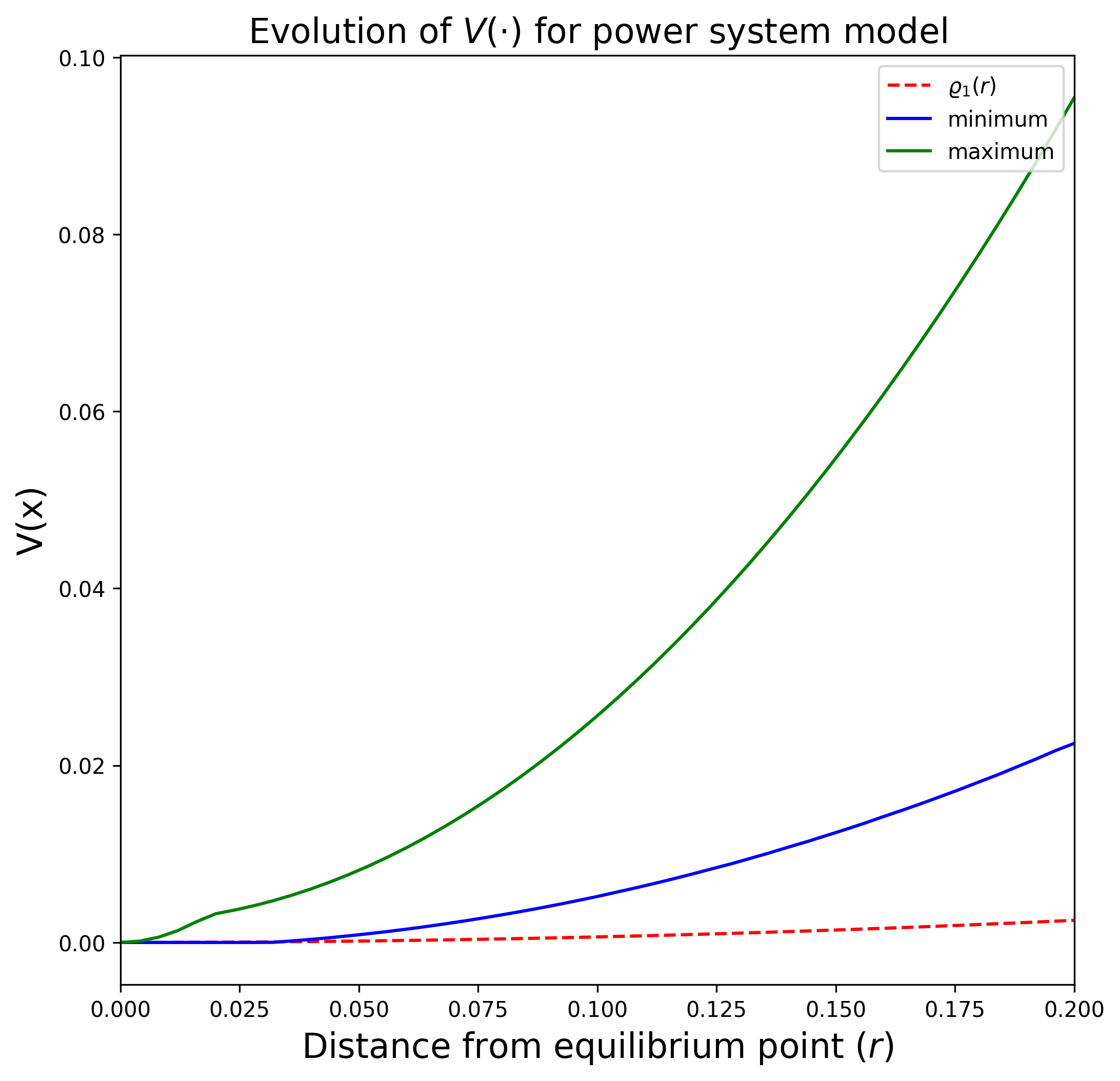}
  \caption{The minimum, on each sphere of radius \(r\), of the Lyapunov function extracted from our algorithm, lies above \(\lowerBound(\cdot)\).}
  \label{fig:power_cand}
\end{subfigure}
\begin{subfigure}{.49\textwidth}
  \centering
  \includegraphics[width=\linewidth]{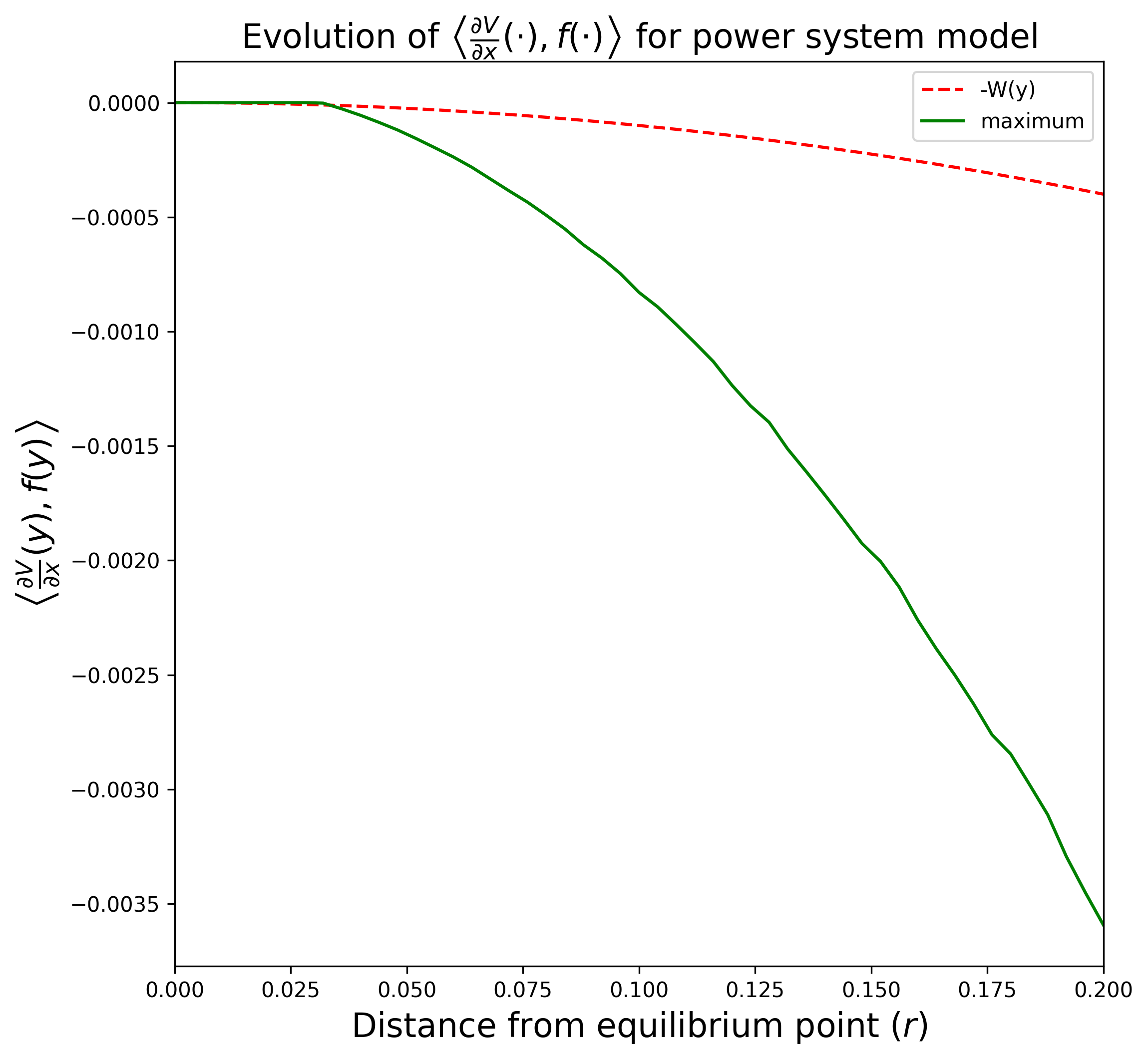}
  \caption{The maximum of \(\inprod{\pdv{\lyapfn}{x}(\cdot)}{\vecfld(\cdot)}\) lies below \(-\stabilityMargin(\cdot)\) on each sphere of radius \(r\).}
  \label{fig:power_grad}
\end{subfigure}
\caption{Illustration of constraint satisfaction corresponding to \eqref{eq:power_model}.}
\label{fig: power}
\end{figure}

\begin{figure}[tbh]
        \centering
        \includegraphics[width=0.6\linewidth]{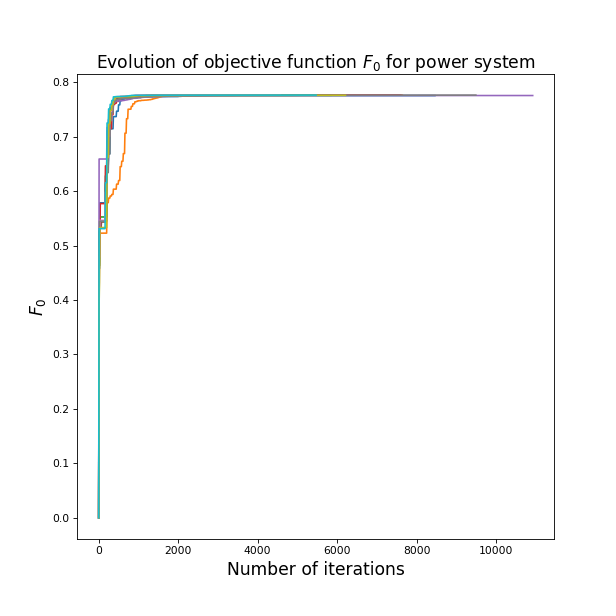}
		\caption{Illustration of the evolution of objective function \(\objective\) for \eqref{eq:power_model} against the number of iterations over 10 executions of the simulated annealing algorithm; the objective $\objective$ converges to the maximum \(0.775\) in 100\% of the executions.}
        \label{fig: power-converge}
\end{figure}

\subsection{Flexibility of basis}

We present a final example illustrating an interesting situation in which the vector field is defined by means of a case statement. This example is lifted from \cite[p. 68]{ref:BhaSze-70}, and the dynamics is given by
\begin{equation}
  \label{eq:bhatia's}
  \begin{aligned}
	  \dot\state_{1} & =
	  \begin{cases}
		  \state_1 & \text{if }\state_{1}^2\state_{2}^2 \geq 1,\\
				2\state_{1}^3\state_{2}^2 - \state_{1} & \text{if } \state_{1}^2\state_{2}^2 < 1,
	\end{cases}\\
	\dot\state_{2} & = -\state_{2}.
\end{aligned}
\end{equation}
Notice that the vector field \eqref{eq:bhatia's} is continuous, and the neighborhood \(\nbhd\) selected above contains both the regions differentiated by the case statement.

\subsubsection*{Candidate Lyapunov triplet for our experiment}
    Our selections were as follows:
\begin{description}
	\item[\ref{d:nbhd}]  \(\nbhd  \Let \set[\big]{ (\state_1, \state_2) \in \R[2] \suchthat \state_1, \state_2 \in [-4, 4] }\).
	\item[\ref{d:pdf}] \(\lowerBound, \marginBound \in \classK\) : \(\lowerBound\left(\radius\right) = 0\) and \(\marginBound\left(\radius\right) =  \frac{\radius^2}{2048}\) for \(\radius \ge 0\).
	\item[\ref{d:dictionary}] The dictionaries were selected to be:
		\begin{align*}
			\basisDict & \Let \set[\big]{ x_1^{i_1} \cdot x_2^{i_2} \suchthat (i_1, i_2) \in\Nz[2], 2 \leq i_1 + i_2  \leq 6 }, \\
				\marginDict & \Let \set[\big]{ x_1^{i_1} \cdot x_2^{i_2}  \suchthat (i_1, i_2) \in\Nz[2], i_1 + i_2 = 2, 4}.
		\end{align*}
\end{description}
In contrast to the rational polynomial functions mentioned in \cite{ref:BhaSze-70} corresponding to this example, we obtained a Lyapunov function within the class of polynomial functions itself without extending it to include rational functions. The outcome of our numerical experiment was the Lyapunov function
\[
	\state\mapsto\lyapfn(\state) =0.003\state_1^2 + 0.069\state_1\state_2 + 1.063\state_1^2.
\]

\begin{figure}[tbh]
        \centering
        \includegraphics[scale = 0.6]{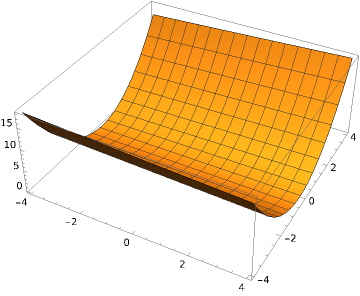} 
		\caption{Lyapunov function to demonstrate the flexibility of the choice of dictionaries for \eqref{eq:bhatia's}.}
        \label{fig: bhatia_lyapunov}
\end{figure}

\begin{figure}[tbh]
\centering
\begin{subfigure}{.49\textwidth}
  \centering
  \includegraphics[width=\linewidth]{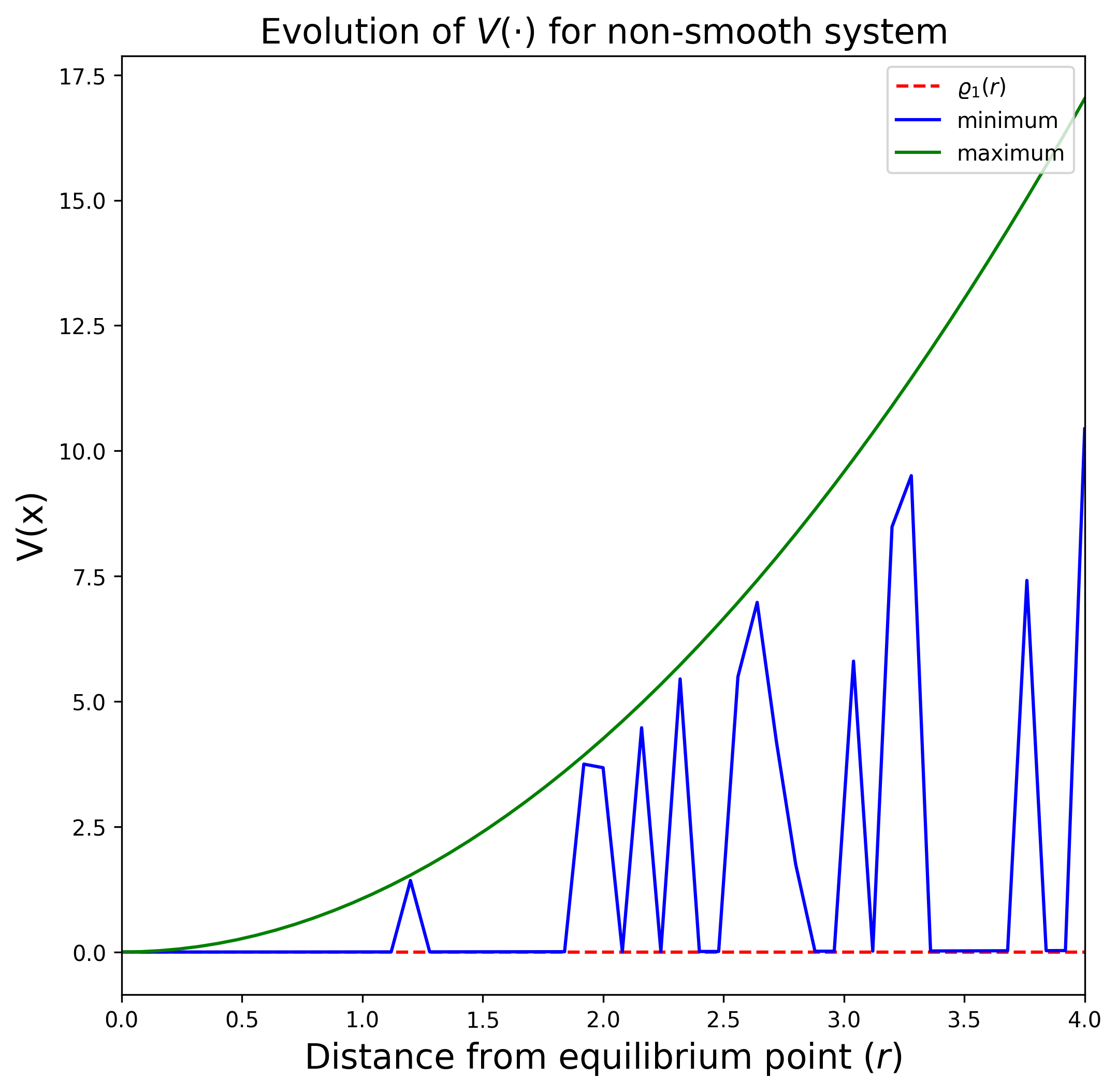}
  \caption{The minimum, on each sphere of radius \(r\), of the Lyapunov function extracted from our algorithm, lies above \(\lowerBound(\cdot)\).}
  \label{fig:bhatia_cand}
\end{subfigure}
\begin{subfigure}{.49\textwidth}
  \centering
  \includegraphics[width=\linewidth]{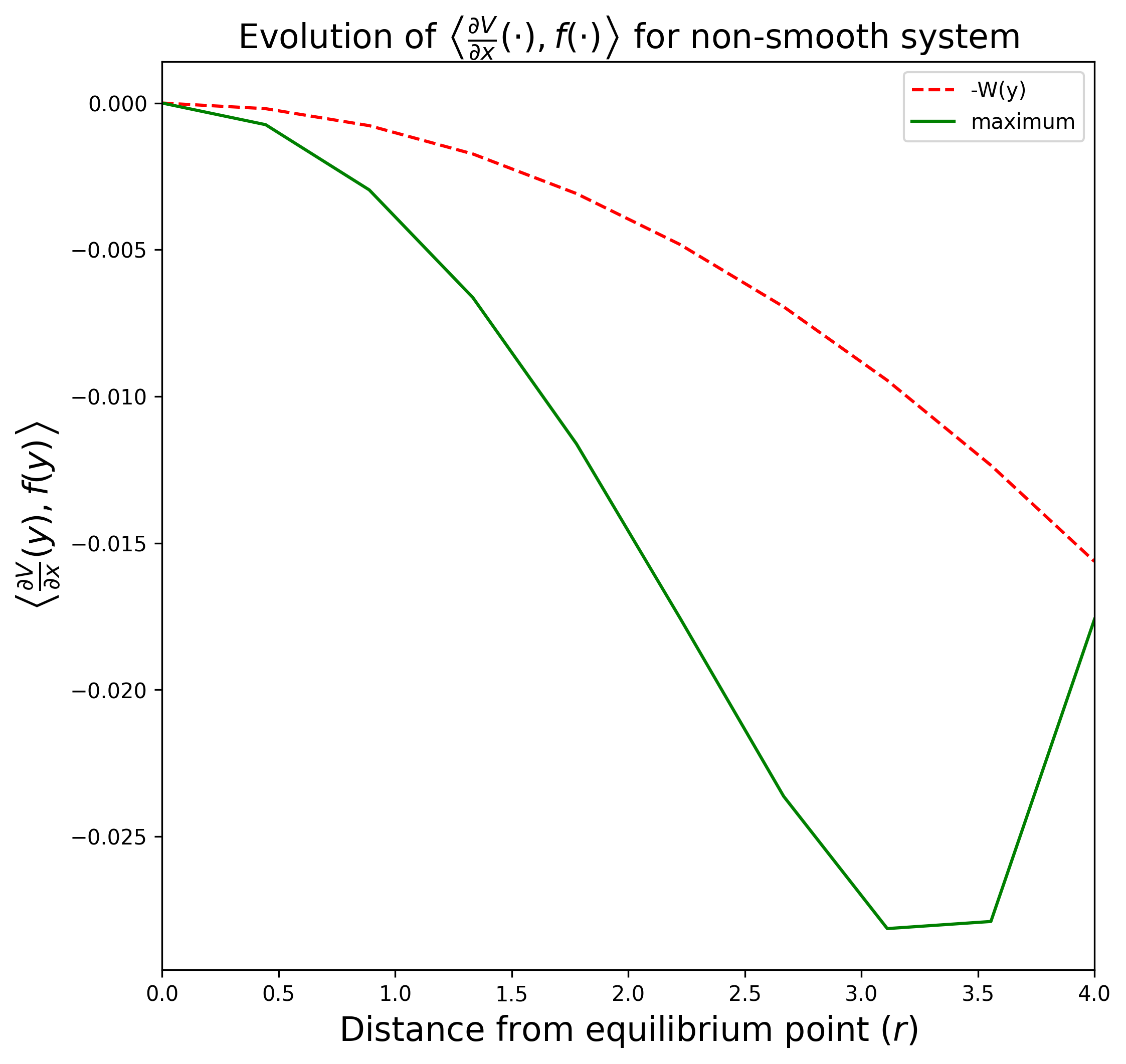}
  \caption{The maximum of \(\inprod{\pdv{\lyapfn}{x}(\cdot)}{\vecfld(\cdot)}\) lies below \(-\stabilityMargin(\cdot)\) on each sphere of radius \(r\).}
  \label{fig:bhatia_grad}
\end{subfigure}
	\caption{Illustration of constraint satisfaction corresponding to \eqref{eq:bhatia's}.}
\label{fig: bhatia}
\end{figure}

\begin{figure}[tbh]
        \centering
        \includegraphics[width=0.6\linewidth]{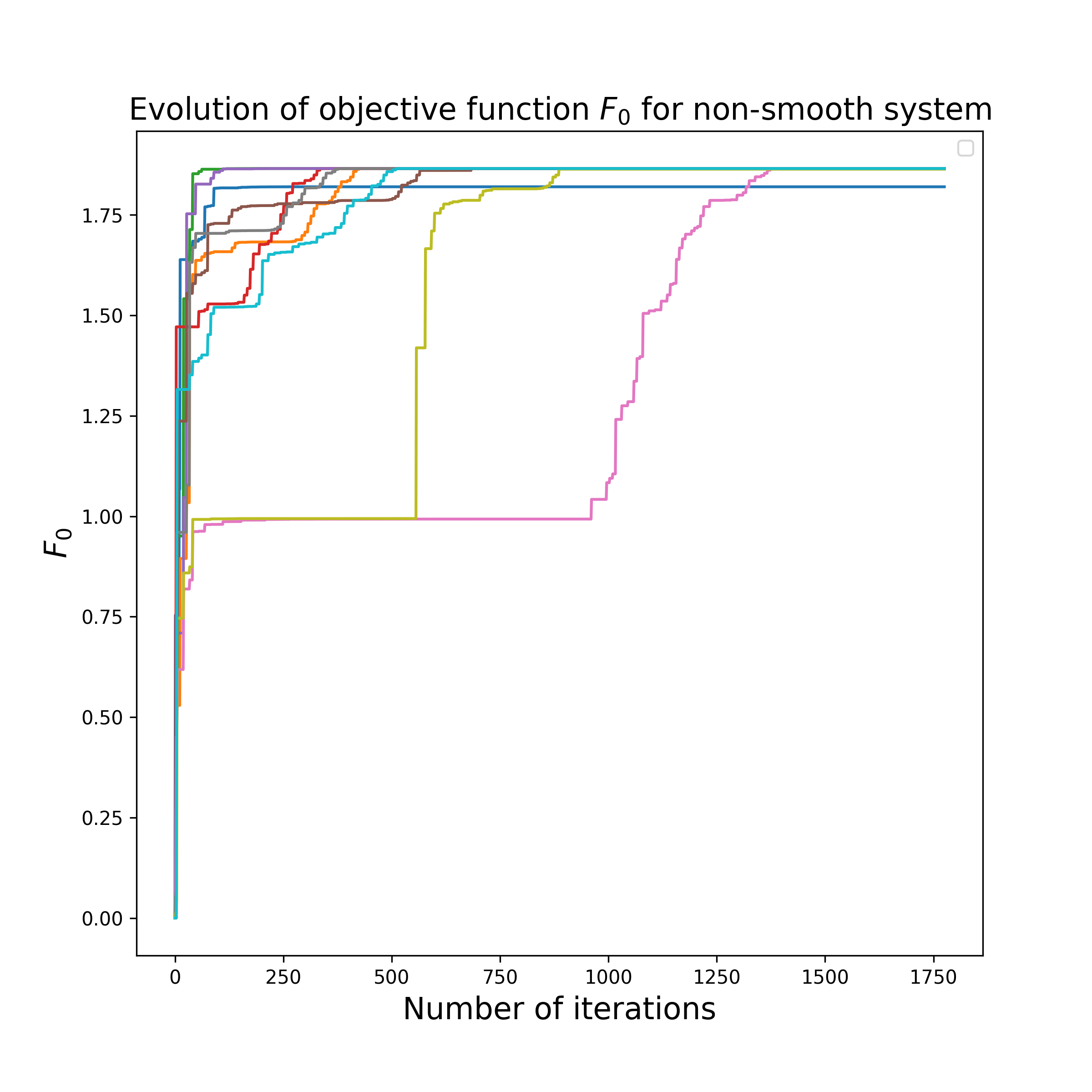}
		\caption{Illustration of the evolution of objective function \(\objective\) for \eqref{eq:bhatia's} against the number of iterations over 10 executions of the simulated annealing algorithm; the objective $\objective$ converges to the maximum \(1.865\) in 90\% of the executions.}
        \label{fig: bhatia-converge}
\end{figure}

\subsection{Rayleigh problem with control constraints}

In this example we address stability of an equilibrium point under \emph{continuous-time model predictive control}. It is well-known that discrete-time model predictive control (MPC) \cite{ref:XiLi-19} is a very powerful constrained control technique, and it consists of the following steps for a system of the form \(x^+ = f(x, u)\):
\begin{itemize}
	\item solve a constrained finite horizon optimal control problem for the given control system with the current state \(x\) as the initial condition, and extract the corresponding sequence of optimal control actions over the horizon,
	\item apply the entry corresponding to initial stage to obtain the next state \(x^+\), 
	\item increment the time by \(1\), and return back to the first step with \(x^+\) as the current state.
\end{itemize}
No analytical expression of the optimal control sequence is available, in general, due to the presence of constraints in the problem, but it is clear that the optimal control actions depend on the initial state \(x\); if the optimal control action is unique at the initial time, then it is a \emph{mapping} of the initial state.\footnote{This feature is exploited by the so-called \emph{explicit MPC} technique to numerically construct this implicit feedback.}

The idea of continuous-time MPC for a control system \(\dot x = f(x) + g(x) u\) follows the same route as outlined above, starting from a continuous-time finite horizon constrained control problem (given an initial condition) over control trajectories, and extracting the optimal control trajectory (assuming uniqueness). This trajectory parametrically depends on the initial condition, and therefore, the optimal control action \(u\opt(0)\) at the initial time \(0\) depends on the initial state \(x\); one denotes this dependence as \(u\opt(0; x)\). The closed-loop system under a continuous-time MPC controller, consequently, is \(\dot x = f(x) + g(x) u\opt(0; x)\); it is immediately clear that \(u\opt(0; \cdot)\) is a feedback. Solving continuous-time constrained optimal control problems is typically time consuming, and it is an extremely challenging matter to evaluate \(u\opt(0; x)\) instantaneously for the continuous-time model. However, if such an implementaion were possible, then \(u\opt(0; \cdot)\) would be evaluated by an oracle and, in general, would not admit an explicit formula.

This example studies stability of an equilibrium point of the nonlinear control system 
\[
	\pmat{\dot{x}_1(t)\\ \dot{x}_2(t)} = \pmat{x_2(t) \\ -x_1(t) + x_2(t)\bigl(1.40 - 0.14 x_2(t)^2\bigr)} + \pmat{0\\ 4} u(t)
\]
under a particular continuous-time MPC strategy; this system features in the benchmark Rayleigh's problem \cite[Example 4.6]{ref:Bet-10}, and for our purposes it will be accompanied by control constraints over a finite horizon. Note that the vector field of the nonlinear system vanishes at the triplet \((x_1, x_2, u) = (0, 0, 0)\). Our intention is to stabilize the origin \((x_1, x_2) = (0, 0)\) under continuous-time MPC in closed-loop, and with this in mind, the following underlying continuous-time optimal control problem for MPC was formulated:
\begin{equation}
\label{eq:quito}
\begin{aligned}
	\minimize_{u:\lcrc{0}{\tfin}\to\R[]} \quad & \int_{0}^{\tfin} \bigl(u(t)^2 + x_1(t)^2\bigr) \, \odif{t} \\
\sbjto \quad &
\begin{cases}
\dot{x}_1(t) = x_2(t), \\
\dot{x}_2(t) = -x_1(t) + x_2(t)\bigl(1.40 - 0.14 x_2(t)^2\bigr) + 4u(t),\\
	\bigl(x_1(0), x_2(0)\bigr) = (a_1, a_2) \text{ (given)}, \\
	\bigl(x_1(\tfin), x_2(\tfin)\bigr) = (0, 0), \\
	|u(t)| \leq 1.
\end{cases}
\end{aligned}
\end{equation}
We picked the horizon \(\tfin = 4.5\). The optimal trajectory construction algorithm QuITO (Quasi Interpolation based Trajectory Optimization) developed in \cite{ref:Ganguly2022} was employed to solve the optimal control problem \eqref{eq:quito} and obtain optimal control trajectories \(\lcrc{0}{\tfin}\ni t\mapsto u\opt\bigl(t; (a_1, a_2)\bigr)\in\R[]\) for each initial condition \((a_1, a_2)\) sufficiently close to \((0, 0)\). The step size for the uniform cardinal grid on \(\lcrc{0}{\tfin}\) was picked to be \(h = 0.09\), and the interpolating Schwarz function was chosen to be \(\R[]\ni t\mapsto \psi(t) = \frac{1}{\sqrt{\pi\mathcal D}} \epower{-\frac{t^2}{h^2 \mathcal D}}\) with the shape parameter \(\mathcal D = 5\), both of which were input parameters to QuITO.

The resulting closed-loop control system was, naturally,
\begin{equation}
	\label{e:clquito}
	\pmat{\dot{x}_1(t)\\ \dot{x}_2(t)} = \pmat{x_2(t) \\ -x_1(t) + x_2(t)\bigl(1.40 - 0.14 x_2(t)^2\bigr)} + \pmat{0\\ 4} u\opt\bigl(0; x_1(t), x_2(t)\bigr).
\end{equation}
We highlight that the right-hand side of \eqref{e:clquito} lacks an analytical expression although off-the-shelf results may be employed to check the mapping \(x\mapsto u\opt(0; x)\) is continuous and it evaluates to \(0\) at \(x = (0, 0)\). This means \((0, 0)\) is an equilibrium point of the preceding closed-loop system; let us check whether it is asymptotically stable.

\subsubsection*{Candidate Lyapunov triplet for our experiment}
Our selections were as follows:
\begin{description}
	\item[\ref{d:nbhd}]  \(\nbhd  \Let \set[\big]{ (\state_1, \state_2) \in \R[2] \suchthat \state_1^2 + \state_2^2  \leq 0.2025}\).
	\item[\ref{d:pdf}] \(\lowerBound, \marginBound \in \classK\) were picked as \(\lowerBound\left(\radius\right) = \frac{\radius^2}{4}\) and \(\marginBound\left(\radius\right) =  \frac{\radius^2}{32}\) for \(\radius \ge 0\).
	\item[\ref{d:dictionary}] The dictionaries were selected to be:
		\begin{align*}
			\basisDict & \Let \set[\big]{ x_1^{i_1} \cdot x_2^{i_2} \suchthat (i_1, i_2) \in\Nz[2], i_1 + i_2  = 2 }, \\ 
			\marginDict & \Let \set[\big]{ x_1^{i_1} \cdot x_2^{i_2} \suchthat (i_1, i_2) \in\Nz[2], i_1 + i_2  = 2, 4, 6 }.
		\end{align*}
\end{description}
The Lyapunov function obtained from our numerical procedure was
\[
	\state\mapsto\lyapfn(\state) = 1.216\state_1^2 + 0.688\state_1\state_2+ 0.948\state_2^2,
\]
which shows that the equilibrium point \((0, 0)\) is indeed asymptotically stable.

\begin{figure}[tbh]
        \centering
        \includegraphics[scale = 0.6]{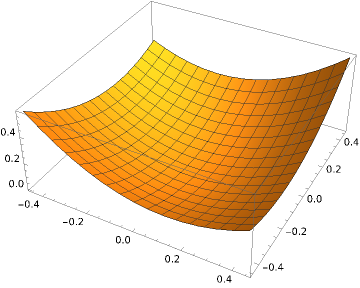} 
		\caption{Lyapunov function for Rayleigh problem in \eqref{eq:quito}.}
        \label{fig: quito_lyapunov}
\end{figure}

\begin{figure}[tbh]
\centering
\begin{subfigure}{.49\textwidth}
  \centering
  \includegraphics[width=\linewidth]{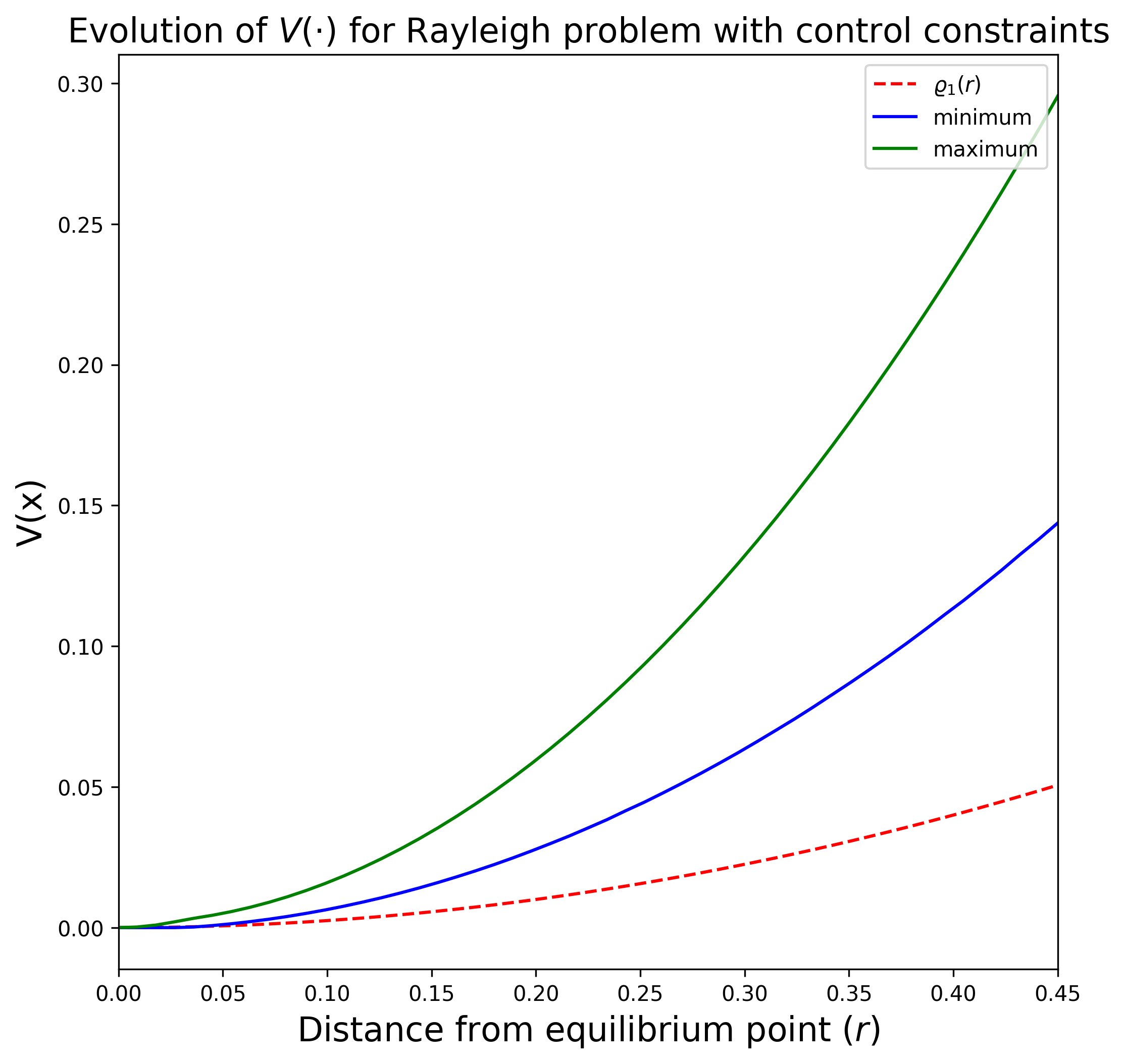}
  \caption{The minimum, on each sphere of radius \(r\), of the Lyapunov function extracted from our algorithm, lies above \(\lowerBound(\cdot)\).}
  \label{fig:quito_cand}
\end{subfigure}
\begin{subfigure}{.49\textwidth}
  \centering
  \includegraphics[width=\linewidth]{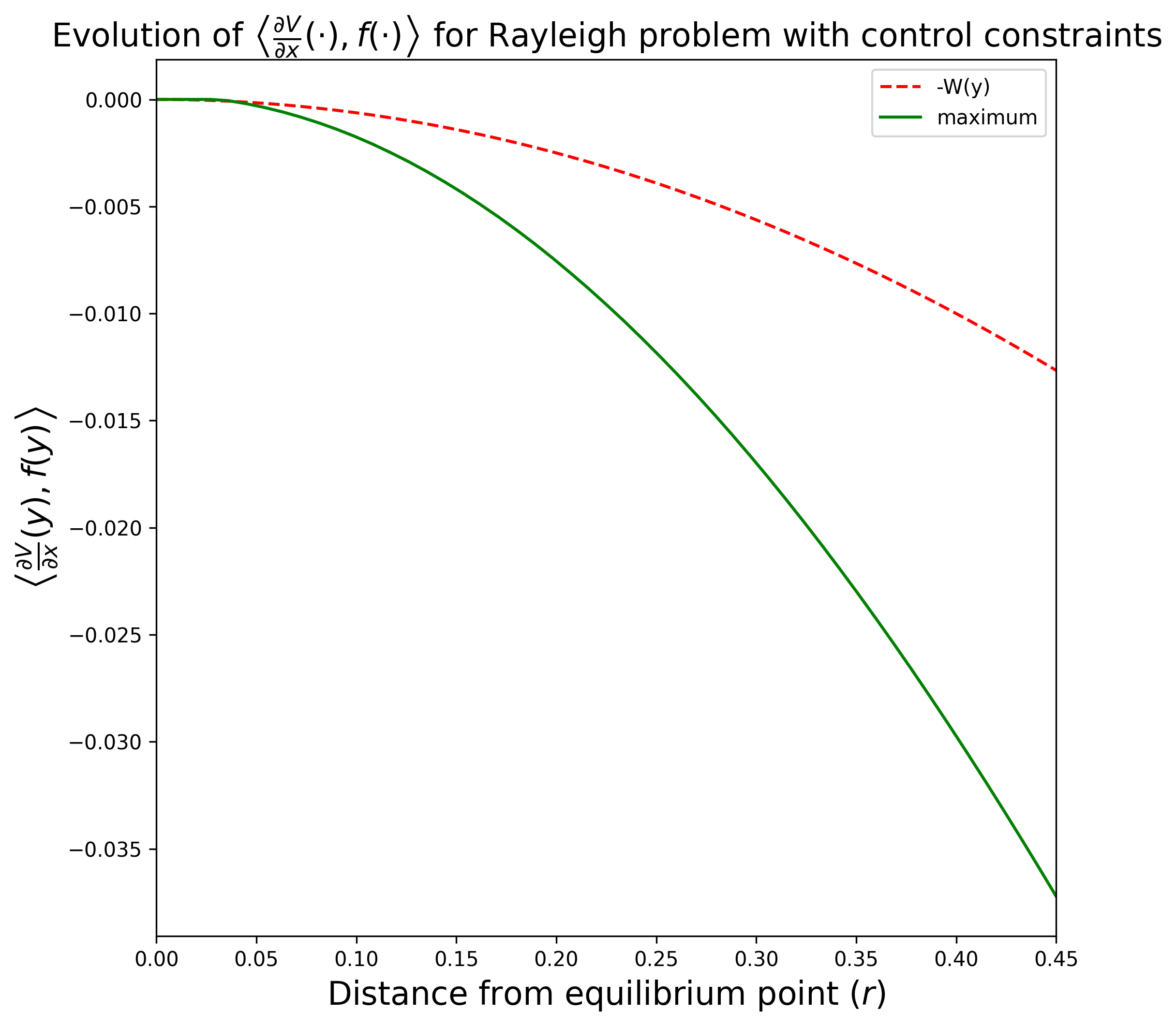}
  \caption{The maximum of \(\inprod{\pdv{\lyapfn}{x}(\cdot)}{\vecfld(\cdot)}\) lies below \(-\stabilityMargin(\cdot)\) on each sphere of radius \(r\).}
  \label{fig:quito_grad}
\end{subfigure}
	\caption{Illustration of constraint satisfaction corresponding to \eqref{eq:quito}.}
\label{fig: quito}
\end{figure}
\begin{figure}[tbh]
        \centering
        \includegraphics[width=0.6\linewidth]{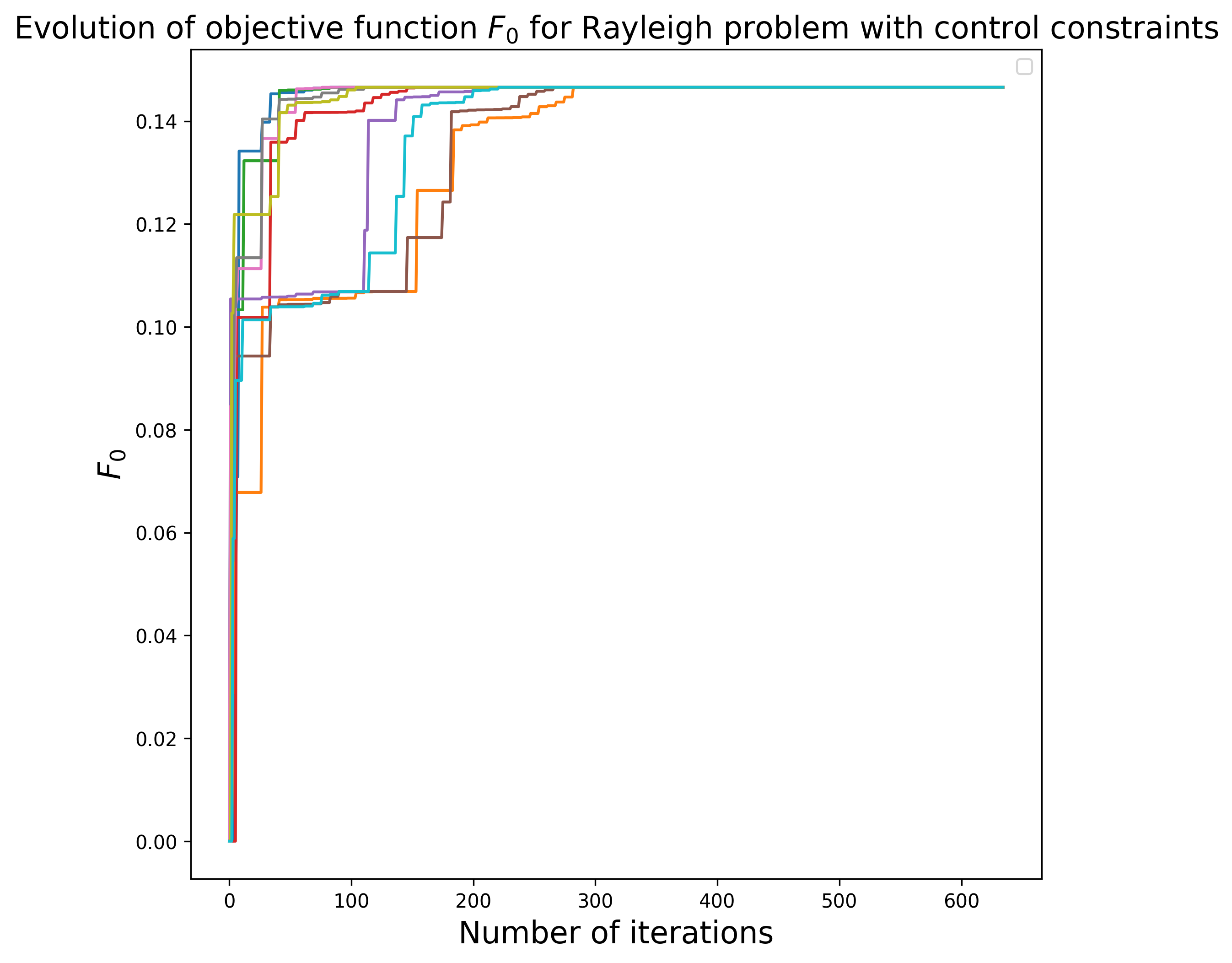}
		\caption{Illustration of the evolution of objective function \(\objective\) for \eqref{eq:quito} against the number of iterations over 10 executions of the simulated annealing algorithm; the objective $\objective$ converges to the maximum \(0.1466\) in 100\% of the executions.}
        \label{fig: quito-converge}
\end{figure}

\section{A quick excursion into instability via Chetaev's Theorem}

In this brief section, we transcend beyond the stability of isolated equilibria into their instability. This gives us one more opportunity to leverage the generality and adaptability of our algorithmic procedure. We consider the problem of determining the instability of an equilibrium point by employing Chetaev's theorem \cite[Chapter V, p.\ 188]{ref:Vid-02}. We refer to the example of spinning of a rigid body provided in \cite[pp. 188-190]{ref:Vid-02}, and express the dynamics as:
\begin{equation}
  \label{eq:chetaev's}
  \begin{aligned}
	  \dot\state_{1} & = a\state_{2}\state_{3},\\
	  \dot\state_{2} & = -b\state_{1}\state_{3},\\
	  \dot\state_{3} & = c\state_{1}\state_{2},
    \end{aligned}
\end{equation}
where for the sake of simplicity, we pick \(a = b = c = 1\). We pick an equilibrium point of the above system of the form \( (0, y_0, 0)\) with \(y_0 \ge 0\) to be specific and transform coordinates as shown in \cite[Example 105, p.\ 189]{ref:Vid-02}.

\subsubsection*{Candidate Lyapunov triplet for our experiment}
    Our selections were as follows:
\begin{description}
	\item[\ref{d:nbhd}]  \(\nbhd  \Let \set[\big]{ (\state_1, \state_2, \state_3) \in \R[3] \suchthat \state_1, \state_2, \state_3 \in [0, 1] }\).
	\item[\ref{d:pdf}] \(\lowerBound \in \classK\): \(\lowerBound\left(\radius\right) = 0\) for \(\radius \ge 0\).
	\item[\ref{d:dictionary}] The dictionary for candidate Lyapunov functions was selected to be
		\begin{align*}
			\basisDict & \Let \set[\big]{ x_1^{i_1} \cdot x_2^{i_2} \cdot x_3^{i_3} \suchthat (i_1, i_2, i_3) \in\Nz[3], 2 \leq i_1 + i_2 + i_3 \leq 6 }.
		\end{align*}
\end{description}
No stability margin was incorporated in the numerical experiment, effectively considering it to be zero. Upon executing the dynamics, the resulting Lyapunov-like function is obtained as \(\state\mapsto \state_{1}^2 + \state_{2}^2 + \state_{3}^2 + \state_1\state_3\), which fulfills the constraints specified in Chetaev's theorem \cite[Theorem 99, p.\ 188]{ref:Vid-02}. This confirms the instability of the system's equilibrium point at the origin \((0, 0, 0)\).

\section{Concluding remarks}
\label{s:concl}

We have given an exposition of a numerically tractable technique for finding Lyapunov functions to test two types of stability properties of equilibria of continuous finite-dimensional vector fields. These two properties are representational, and our technique carries over to other qualitative properties of equilibria that are captured (equivalently) by means of appropriate pointwise behavior of Lyapunov-like functions. Indeed, the technique readily applies to robust and stochastic versions of stability and, as such, will be useful for verifying input-to-state stability (ISS) and its allied versions in the robust setting, as well as for verifying different types of stochastic stability. Extensions of our results to the synthesis of control Lyapunov functions can also be readily carried out by means of our techniques. Results along some of the aforementioned directions will be reported subsequently.



\begin{thebibliography}{MWY{\etalchar{+}}20}

\bibitem[AMP13]{ref:AngMilPap-13}
M.~Anghel, F.~Milano, and A.~Papachristodoulou, \emph{Algorithmic construction
  of {L}yapunov functions for power system stability analysis}, IEEE
  Transactions on Circuits and Systems I: Regular Papers \textbf{60} (2013),
  no.~9, 2533--2546, doi: \url{https://doi.org/10.1109/TCSI.2013.2246233}.

\bibitem[AP15]{ref:AndPap-15}
J.~Anderson and A.~Papachristodoulou, \emph{Advances in computational
  {L}yapunov analysis using sum-of-squares programming}, Discrete and
  Continuous Dynamical Systems. Series B. \textbf{20} (2015), no.~8,
  2361--2381, doi: \url{https://doi.org/10.3934/dcdsb.2015.20.2361}.

\bibitem[Ber18]{ref:Ber-18}
D.~S. Bernstein, \emph{Scalar, {V}ector, and {M}atrix {M}athematics}, Princeton
  University Press, Princeton, NJ, 2018, Theory, facts, and formulas, Revised
  and expanded edition.

\bibitem[Bet10]{ref:Bet-10}
J.~T. Betts, \emph{Practical {M}ethods for {O}ptimal {C}ontrol and estimation
  using {N}onlinear {P}rogramming}, 2nd ed., Advances in Design and Control,
  vol.~19, SIAM, Philadelphia, PA, 2010, doi:
  \url{https://doi.org/10.1137/1.9780898718577}.

\bibitem[BS02]{ref:BhaSze-70}
N.~P. Bhatia and G.~P. Szeg\H{o}, \emph{Stability {T}heory of {D}ynamical
  {S}ystems}, Classics in Mathematics, Springer-Verlag, Berlin, 2002, Reprint
  of the 1970 original.

\bibitem[CG18]{ref:CamGar-18}
M.~C. Campi and S.~Garatti, \emph{Introduction to the {S}cenario {A}pproach},
  MOS-SIAM Series on Optimization, vol.~26, Society for Industrial and Applied
  Mathematics (SIAM), 2018, doi:
  \url{https://doi.org/10.1137/1.9781611975444.ch1}.

\bibitem[DACC22]{ref:DasAraCheCha-22}
S.~Das, A.~Aravind, A.~Cherukuri, and D.~Chatterjee, \emph{Near-optimal
  solutions of convex semi-infinite programs by targeted sampling}, Annals of
  Operations Research \textbf{318} (2022), 129--146, doi:
  \url{https://doi.org/10.1007/s10479-022-04810-4}.

\bibitem[GRCB22]{ref:Ganguly2022}
S.~Ganguly, N.~Randad, D.~Chatterjee, and R.~Banavar, \emph{Constrained
  trajectory synthesis via quasi-interpolation}, 61st IEEE Conference on
  Decision and Control (CDC), 2022, pp.~4533--4538.

\bibitem[Hah67]{ref:Hah-67}
W.~Hahn, \emph{Stability of {M}otion}, Springer-Verlag New York, Inc., New
  York,, 1967, Translated from the German manuscript by Arne P. Baartz.

\bibitem[HCL14]{ref:HanCheLuk-14}
D.~Han, G.~Chesi, and C.~K. Luk, \emph{Homogeneous polynomial {L}yapunov
  functions for robust local synchronisation with time-varying uncertainties},
  IET Control Theory \& Applications \textbf{8} (2014), no.~10, 855--862, doi:
  \url{https://doi.org/10.1049/iet-cta.2013.0742}.

\bibitem[JP23]{ref:JonPee-23}
M.~Jones and M.~M. Peet, \emph{A converse sum of squares {L}yapunov function
  for outer approximation of minimal attractor sets of nonlinear systems},
  Journal of Computational Dynamics \textbf{10} (2023), no.~1, 48--74, doi:
  \url{https://doi.org/10.3934/jcd.2022019}.

\bibitem[KA15]{ref:KunAng-15}
S.~Kundu and M.~Anghel, \emph{A sum-of-squares approach to the stability and
  control of interconnected systems using vector {L}yapunov functions}, 2015
  American Control Conference (ACC), 2015, doi:
  \url{https://doi.org/10.1109/ACC.2015.7172121}, pp.~5022--5028.

\bibitem[KSCN22]{ref:KumSriChaNag-22}
Y.~Kumar, S.~Srikant, D.~Chatterjee, and M.~Nagaraha, \emph{Sparse optimal
  control problems with intermediate constraints: necessary conditions},
  Optimal Control Applications \& Methods \textbf{43} (2022), no.~2, 369--385,
  doi: \url{https://doi.org/10.1002/oca.2807}.

\bibitem[Lan93]{ref:Lan-93}
S.~Lang, \emph{Real and {F}unctional {A}nalysis}, 3 ed., Graduate Texts in
  Mathematics, vol. 142, Springer-Verlag, New York, 1993, doi:
  \url{https://doi.org/10.1007/978-1-4612-0897-6}.

\bibitem[Lib12]{ref:Lib-12}
D.~Liberzon, \emph{Calculus of {V}ariations and {O}ptimal {C}ontrol {T}heory},
  Princeton University Press, Princeton, NJ, 2012, A concise introduction.

\bibitem[MCB20]{ref:MisChaBan-20}
P.~K. {Mishal Assif}, D.~Chatterjee, and R.~Banavar, \emph{Scenario approach
  for minmax optimization with emphasis on the nonconvex case: Positive results
  and caveats}, SIAM Journal on Optimization (2020), 1119–1143, doi:
  \url{https://doi.org/10.1137/19M1271026}.

\bibitem[MWY{\etalchar{+}}20]{ref:MenWanYanXieGuo-20}
F.~Meng, D.~Wang, P.~Yang, G.~Xie, and F.~Guo, \emph{Application of
  sum-of-squares method in estimation of region of attraction for nonlinear
  polynomial systems}, IEEE Access \textbf{8} (2020), 14234--14243, doi:
  \url{https://doi.org/10.1109/ACCESS.2020.2966566}.

\bibitem[Niu21]{ref:Niu2021}
H.~Niu, \emph{Analysis and stability control of a novel {5D} hyperchaotic
  system}, Journal of Robotics, Networking and Artificial Life \textbf{8}
  (2021), 245--248, doi: \url{https://doi.org/10.2991/jrnal.k.211108.003}.

\bibitem[PAV{\etalchar{+}}21]{ref:SOSTOOLS}
A.~Papachristodoulou, J.~Anderson, G.~Valmorbida, S.~Prajna, P.~Seiler, P.~A.
  Parrilo, M.~M. Peet, and D.~Jagt, \emph{{SOSTOOLS}: Sum of squares
  optimization toolbox for {MATLAB}}, doi:
  \url{http://arxiv.org/abs/1310.4716}, 2021, Available from
  \url{https://github.com/oxfordcontrol/SOSTOOLS}.

\bibitem[PC23]{ref:ParCha-23}
P.~Paruchuri and D.~Chatterjee, \emph{Attaining the {C}hebyshev bound in
  optimal learning: a numerical algorithm}, arXiv preprint; doi:
  \url{https://doi.org/10.48550/arXiv.2307.01304}, 2023.

\bibitem[PP02]{ref:PapPra-02}
A.~Papachristodoulou and S.~Prajna, \emph{On the construction of {L}yapunov
  functions using the sum of squares decomposition}, Proceedings of the 41st
  IEEE Conference on Decision and Control, vol.~3, 2002, doi:
  \url{https://doi.org/10.1109/CDC.2002.1184414}, pp.~3482--3487.

\bibitem[PPSP05]{ref:PraPapSelPar-05}
S.~Prajna, A.~Papachristodoulou, P.~Seiler, and P.~A. Parrilo, \emph{S{OSTOOLS}
  and its control applications}, Positive {P}olynomials in {C}ontrol, Lecture
  Notes in Control and Information Sciences, vol. 312, Springer, Berlin, 2005,
  doi: \url{https://doi.org/10.1007/10997703_14}, pp.~273--292.

\bibitem[PS22]{ref:PolSze-22}
P.~Polcz and G.~Szederk\'{e}nyi, \emph{Lyapunov function computation for
  autonomous systems with complex dynamic behavior}, European Journal of
  Control \textbf{65} (2022), Paper No. 100619, 14, doi:
  \url{https://doi.org/10.1016/j.ejcon.2022.100619}.

\bibitem[Son93]{ref:Son-93}
E.~D. Sontag, \emph{Neural networks for control}, Essays on control:
  perspectives in the theory and its applications ({G}roningen, 1993), Progress
  in Systems and Control Theory, vol.~14, Birkh\"{a}user Boston, Boston, MA,
  1993, pp.~339--380.

\bibitem[SS97]{ref:SonSus-97}
E.~D. Sontag and H.~J. Sussmann, \emph{Complete controllability of
  continuous-time recurrent neural networks}, Systems \& Control Letters
  \textbf{30} (1997), no.~4, 177--183, doi:
  \url{https://doi.org/10.1016/S0167-6911(97)00002-9}.

\bibitem[TP08]{ref:TanPac-08}
W.~Tan and A.~Packard, \emph{Stability region analysis using polynomial and
  composite polynomial {L}yapunov functions and sum-of-squares programming},
  IEEE Transactions on Automatic Control \textbf{53} (2008), no.~2, 565--571,
  doi: \url{https://doi.org/10.1109/TAC.2007.914221}.

\bibitem[Vid02]{ref:Vid-02}
M.~Vidyasagar, \emph{Nonlinear {S}ystems {A}nalysis}, Classics in Applied
  Mathematics, vol.~42, Society for Industrial and Applied Mathematics (SIAM),
  Philadelphia, PA, 2002, Reprint of the second (1993) edition. doi:
  \url{https://doi.org/10.1137/1.9780898719185}.

\bibitem[VV85]{ref:VanVid-85}
A.~Vannelli and M.~Vidyasagar, \emph{Maximal {L}yapunov functions and domains
  of attraction for autonomous nonlinear systems}, Automatica \textbf{21}
  (1985), no.~1, 69--80, doi:
  \url{https://doi.org/10.1016/0005-1098(85)90099-8}.

\bibitem[XL19]{ref:XiLi-19}
Y.~Xi and D.~Li, \emph{Predictive control}, John Wiley \& Sons, 2019.

\bibitem[ZSWX23]{ref:ZhaSonWanXue-23}
S.~Zhang, S.~Song, L.~Wang, and B.~Xue, \emph{Stability verification for
  heterogeneous complex networks via iterative {SOS} programming}, IEEE Control
  Systems Letters \textbf{7} (2023), 559--564, doi:
  \url{https://doi.org/10.1109/LCSYS.2022.3202826}.

\end{thebibliography}

\newcommand{\etalchar}[1]{$^{#1}$}
\providecommand{\bysame}{\leavevmode\hbox to3em{\hrulefill}\thinspace}
\providecommand{\MR}{\relax\ifhmode\unskip\space\fi MR }
\providecommand{\MRhref}[2]{%
  \href{http://www.ams.org/mathscinet-getitem?mr=#1}{#2}
}
\providecommand{\href}[2]{#2}

\end{document}